\documentclass[12pt]{article}

\usepackage[utf8]{inputenc}
\usepackage{amsmath,amssymb,amsfonts,amsthm}
\usepackage[margin=1in]{geometry}
\usepackage{setspace}
\usepackage[
backend=biber,style=apa,citestyle=authoryear,
maxbibnames=10,minbibnames=1,uniquename=false,uniquelist=false,maxcitenames=3]{biblatex}
\usepackage[hidelinks]{hyperref}
\hypersetup{
    colorlinks=true,
    linkcolor=blue,
    citecolor=blue
}

\usepackage{graphicx}
\usepackage{array}
\usepackage[labelfont=bf]{caption}
\usepackage{subcaption}
\usepackage{float}
\usepackage{dcolumn}
\usepackage{placeins}
\newcolumntype{M}[1]{>{\centering\arraybackslash}m{#1}}
\newcolumntype{L}[1]{>{\arraybackslash}m{#1}}
\newcolumntype{P}[1]{>{\centering\arraybackslash}p{#1}}
\newcolumntype{B}[1]{>{\centering\arraybackslash}b{#1}}
\newcolumntype{d}[1]{>{\centering\arraybackslash}D{.}{.}{#1}}

\newtheorem{theorem}{Theorem}

\theoremstyle{plain}

\newtheorem{assumption}{Assumption}

\newtheorem{corollary}{Corollary}

\newtheorem{example}{Example}

\newtheorem{problem}{Problem}
\newtheorem{proposition}{Proposition}

\theoremstyle{definition}

\newtheorem{definition}{Definition}

\newtheorem{remark}{Remark}[section]

\usepackage{threeparttable, booktabs}
\usepackage{comment}

\begin{document}
\def\sym#1{\ifmmode^{#1}\else\(^{#1}\)\fi}

\title{Collective models and the marriage market}
\author{Simon Weber$^\star$}
\date{\today. }
\maketitle

\begin{abstract}
In this paper, I develop an integrated approach to collective models and matching models of the marriage market. In the collective framework, both household formation and the intra-household allocation of bargaining power are taken as given. This is no longer the case in the present contribution, where both are endogenous to the determination of equilibrium on the marriage market. I characterize a class of ``proper'' collective models which can be embedded into a general matching framework with imperfectly transferable utility. In such models, the bargaining sets are parametrized by an analytical device called \emph{distance function}, which plays a key role both for writing down the usual stability conditions and for estimation. In general, however, distance functions are not known in closed-form. I provide an efficient method for computing distance functions, that works even with the most complex collective models. Finally, I provide a fully-fledged application using PSID data. I identify the sharing rule and its distribution and study the evolution of the sharing rule and housework time sharing in the United States since 1969. In a counterfactual experiment, I simulate the impact of closing the gender wage gap.
\end{abstract}

\bigskip
{\footnotesize
\thanks{\begin{singlespace}$^\star$University of York, \emph{simon.weber@york.ac.uk}. I thank Alfred Galichon for his guidance and support. This paper has greatly benefited from comments from Laurens Cherchye, Edoardo Ciscato, Bram De Rock, Arnaud Dupuy, Marco Franscesconi, Alfred Galichon, Jorge Luis Garc\'{i}a, James Heckman, Sonia Jaffe, Magne Mogstad, Jack Mountjoy, Noemi Nocera, Frederic Vermeulen, Paul Vertier and Alessandra Voena and seminar participants at the Becker Friedman Institute, CEHD, CERGE-EI, KU Leuven, LISER, Royal Holloway, the Toulouse School of Economics, the University of Chicago and the University of York. I am particularly grateful to the Becker Friedman Institute, the CEHD, KU Leuven and the University of Chicago for hosting me while most of this paper was written.\end{singlespace}}
}

\clearpage\newpage

\section{Introduction}

Power matters. That is the key insight that stems from the collective model literature developed since the seminal paper of \textcite{Chiappori1988}. In the collective representation of the household, partners engage in bargaining over which goods to consume or produce, and reach Pareto-efficient decisions. The allocation of power (summarized by ``Pareto weights''\footnote{The usual way of modelling efficiency is that partners maximize a weighted sum of their utilities. These weights, called Pareto weights, are an unknown function of prices, income and power ``shifters'' known as distribution factors. Pareto weights provide an intuitive measure of bargaining power. An alternative, equivalent measure of power are sharing rules, which describe what share of the household resources are effectively received by one partner.}) is therefore critical to understand many important household decisions such as labor market participation, time allocation within the household, public good expenditures, or fertility choices. However, it must be understood that in collective models, the Pareto weights and household formation are taken as given. In this paper, I propose to endogenize both, by merging collective models and matching models of the marriage market into a unique framework. To do so, I rely on the recent matching framework with imperfectly transferable utility (ITU) introduced in \textcite{GalichonKominersWeber2019} (GKW hereafter). My contribution is threefold. First, I characterize a class of collective models, coined ``proper collective models'', that can be embedded in the ITU framework. Second, I develop the computational methods so that even the most complex proper collective models are empirically tractable. Finally, I provide a new estimation strategy and a fully-fledged empirical application.

The idea that the allocation of bargaining power within couples and marriage markets are somehow related is not new. It is implicit in many papers from the collective literature that use sex ratios as a distribution factor\footnote{Distribution factors are shifters that affect bargaining power, but not preferences or the budget set of the household members.}  \parencite{ChiapporiFortinLacroix2002}. Intuitively, a marriage market in which women are more scarce would see a shift of the balance of power in their favor. The same can be said about wage ratios \parencite{BrowningGoertz2012}, which is another example of commonly used distribution factors\footnote{For an extensive list, see \textcite{BrowningChiapporiWeiss2014}, chapter 5. Divorce laws, age and education differences, or relative income are commonly used distribution factors as well.}. Implicitly, it is assumed that as women's wage increase relatively to the wage of men, their outside option on the marriage market (be it staying single or divorcing) improves, and so does their bargaining power. In fact, \textcite{Becker1973} already pushed these ideas forward, by showing that the market structure (e.g. the relative supply of men and women) governs the division of marital surplus between partners. This suggests that there could be significant benefits from bringing together collective models and matching models of the marriage market into a unified framework. Or, to put it in Chiappori's words, ``the next step, obviously, would be an ``upstream'' theory that would \emph{endogenize} both household composition (``who marries whom'') and the resulting intrahousehold allocation of power'' \parencite{Chiappori2017}. This paper provides a way of achieving this goal. Therefore, in my framework, distribution factors have a priori no effect on the intra-household allocation of power. Instead, the way they influence power allocation arises naturally from the model itself.

Several papers have been successful in connecting matching and collective models, although most of them rely on a matching model with transferable utility (TU).
In such models, it is assumed that some technology (money, a numeraire good, etc.) allows partners to transfer utility to each other at a constant rate of exchange. This means the utility functions featured in the underlying collective model are restricted to belong to a certain class \parencite{ChiapporiGugl2014}. On the other hand, we gain the computational tractability of the TU framework, since in that case the optimal matching on the marriage market is the one that maximizes total surplus. This allows us to use a battery of well-known computational techniques from linear programming and optimal transportation \parencite{Galichon2016}. However, it turns out that transferable utility is a demanding assumption. In that case, partners behave as a single decision maker: they maximize total utility (or surplus). Thus, public good consumption does not depend on the allocation of bargaining power. Power, here, does not matter (at least when it comes to explaining public good expenditures made by couples), a finding at odds with the empirical literature on collective models, see e.g. \textcite{AttanasioLechene2002}.

Consequently, it is preferable to consider collective models in which utility is (possibly) imperfectly transferable. This means however, that we need the corresponding, fully fledged, matching framework featuring imperfectly transferable utility. In this paper, I rely on the recent work by \textcite{GalichonKominersWeber2019}. They provide a general matching model with imperfectly utility and heterogeneity in tastes, that includes as particular cases matching with transferable utility, non-transferable utility, matching with taxes, etc. The GKW setting is first and foremost a matching framework. It is assumed that men (or firms) and women (or workers) meet on a frictionless marriage (or labor) market. The absence of frictions (which we retain throughout this paper) is an acceptable assumption in our case, as the focus here is on who marries whom and who gets what within couples. The framework is also completely static: divorce and remarriage are absent from the model. Upon meeting, partners bargain and choose a utility allocation\footnote{Therefore, partners make binding agreements on the marriage market. For an important alternative view with bargaining in marriage, see \textcite{Pollak2019}.} $(u,v)$ that belongs to some bargaining set $\mathcal{F}$. The frontier of the bargaining sets need not be a straight line of slope minus one, as would be the case with transferable utility. Nor these sets need to be convex. These sets must, however, meet a certain number of requirements, in which case they are called ``proper'' bargaining sets. To define equilibrium in the ITU framework, stability is used as the solution concept. In equilibrium, it must not be the case that a married individual would rather be single, or that any two individuals would rather be together than in their respective match (or singlehood state). Writing down stability conditions is not as trivial in the ITU case than it is in the transferable utility case. GKW make use of a specific parametrization of the bargaining set to write down stability conditions, called \emph{distance functions} (see section \ref{sec:preamble} for a reminder on GKW). The distance function is key for most of the results presented in that paper, and also plays an important role in estimating ITU models.

The setting introduced in GKW offers a natural way of embedding collective models into a matching framework with imperfectly transferable utility. Before doing so, however, there are two important issues to solve. First, bargaining sets are primitives in GKW, and they are assumed to be proper. However, it is unclear which class of utility functions (and collective models) will generate proper bargaining sets. Therefore, it is necessary to provide sufficient conditions on the underlying collective model for the resulting bargaining sets to be proper. Second, assuming that bargaining sets are proper, it is crucial to be able to compute distance functions for the model to be estimated. In some cases, the distance function can be derived in closed-form. However, as soon as a collective model features labor supply, public goods, time use, household production, or fertility choices, it will produce a bargaining set whose associated distance function cannot be known in closed-form (except in some trivial cases). This is a serious threat for the empirical tractability of the framework.

In this paper, I aim at addressing these concerns. I characterize collective models that will generate proper bargaining sets. I call these collective models ``proper'' collective models as well, because it is then possible to embed such models into the ITU machinery from GKW. I also provide a simple way of computing distance functions. The method works for any proper bargaining set and therefore is not restricted to be used with collective models, but could be applied to the other examples presented in GKW. In the case where a proper collective models has been used to generate the bargaining sets, the computation method also shows interesting connections with the collective model literature. Finally, I provide a new estimation strategy based on a mathematical program with equilibrium constraints (MPEC) \parencite{SuJudd2012} and a fully-fledged application with non-trivial preferences. This is in stark contrast with GKW's application (see appendix D in that paper), which relied on a deliberately simple collective model. Partners only consumed a private good, and the main appeal of the model was that the resulting bargaining sets interpolated between TU and nontransferable utility (NTU). Moreover, the distance function was known is closed-form in their application, in which case estimating the model poses little computational challenges. Instead, I bring a rich collective model to the data: partners consume a composite private good, enjoy leisure, spend time on the labor market and invest time in housework to produce a public good. I make use of this setting to study the evolution of the sharing rule in the United States since 1969 and of housework time sharing. The application also demonstrates the strength of the approach advocated here. First and foremost, since Pareto weights (or equivalently, sharing rules) are endogenized, we can recover them and know their distribution. I show that the sharing rule (the woman's share of total resources) has increased from $0.392$ in 1969 to $0.435$ in 2017 (on average). Second, we can construct counterfactual experiments. For example, I consider what would happen in 2017 were the gender wage gap to disappear (on average). I show that, in that case, the sharing rule  would increase from $0.435$ to $0.514$ in 2017, and the sharing of housework (the woman's share of total housework time) would decrease from $63\%$ to $55\%$.

\textbf{An overview of the literature.} The starting point of this paper are collective models. An excellent overview of the literature is provided in \textcite{BrowningChiapporiWeiss2014}. The relevance of collective models with respect to the classical unitary framework has been illustrated in several papers, e.g. in \textcite{LundbergPollakWales1997} and \textcite{AttanasioLechene2014} for example. Collective models are particularly suited to a context in which the main source of public consumption are children. In this case, if partners have heterogenous tastes over public goods, a shift in bargaining power may influence the well-being of children. Such models have been studied in e.g. \textcite{BlundellChiapporiMeghir2005} and applied in \textcite{CherchyeDeRockVermeulen2012a}.
Aside from testing the relevance of the collective approach, most of the literature has focused on the identification of the Pareto weights/sharing rules (sometimes only up to a constant). Progress has been made in several directions. \textcite{BrowningBonke2006} relied on direct survey evidence. \textcite{ChiapporiFortinLacroix2002} or \textcite{Couprie2007} use assignable goods (private goods whose consumption is observed for both partners) to identify the sharing rule up to a constant. \textcite{LiseSeitz2011} identify the sharing rule parametrically using a discrete choice model of labor supply. \textcite{BrowningChiapporiLewbel2013} and \textcite{LewbelPendakur2008} recover the sharing rule while estimating nonlinear demand systems. Finally, several papers, e.g. \textcite{CherchyeDeRockLewbelEtAl2015}, have used a revealed preference approach to obtain bounds on the sharing rules.

In this paper, I take a different route by using the marriage market to endogenize Pareto weights. I draw insights from Becker's work \parencite{Becker1973}, in which surplus sharing (utility is perfectly transferable in his paper) within couples depends endogenously on population supplies\footnote{\textcite{Becker1973}: ``Theory does not take the division of output between mates as given, but rather derives it from the nature of the marriage market equilibrium''.}. Despite these early contributions, attempts to reconcile collective models and models of the marriage market with transferable utility are only very recent. The mechanics are illustrated in \textcite{Chiappori2012}, and \textcite{BrowningChiapporiWeiss2014} contains interesting examples. Full applications can be found in \textcite{ChooSeitz2013} and \textcite{ChiapporiCostaDiasMeghir2018}. The search and matching literature has produced several papers that account for both household formation and intra-household bargaining with transferable utility, for example \textcite{GousseJacquemetRobin2017} or \textcite{Calvo2021-hv}.

There have been relatively few attempts to integrate collective models into a general ITU matching framework. \textcite{ColesFrancesconi2019} do so in a search-and-matching framework and provide an application in the simpler NTU case. The main reference on frictionless matching with ITU for the present contribution is \textcite{GalichonKominersWeber2019}, and papers cited therein. An application of the \textcite{GalichonKominersWeber2019} framework can be found in \textcite{GayleShephard2019}. Their characterization of proper collective models is not as general as in this paper however. In addition, their application focuses on optimal taxation questions and contains a limited number of types (three types of men and three types of women on a given market), so computational issues are less severe. It also studies a single period of time, whereas I estimate my model for multiple marriage markets in the US since 1969. Finally, several papers from the revealed preferences literature reconcile collective models and the marriage market, for example \textcite{CherchyeDemuynckDeRockEtAl2017} and \textcite{Cherchye2020-im}. These papers use a revealed preference approach to household consumption and stability on the marriage market to tighten the bounds on the sharing rule. The main advantage of their approach is that it is fully nonparametric ; on the other hand, sharing rules are only set-identified, and it is more difficult to include noneconomic marriage gains in their model and conduct counterfactual experiments.

\textbf{Organization of the paper.} Section \ref{sec:preamble} sets the stage by reminding the reader of the ITU matching framework. Section \ref{sec:bargaining} characterizes proper collective models. In section \ref{sec:compmethod}, I introduce a computation method to solve general ITU models. Section \ref{sec:estimation} provides an estimation method based on a mathematical program with equilibrium constraints. In section \ref{sec:proofofconcept}, I provide an empirical application with a collective model featuring private and public consumption as well as labor supplies. Finally, section \ref{sec:conclusion} concludes. All proofs are placed in appendix \ref{app:proofs}.

\section{Setting the stage}\label{sec:preamble}
\subsection{The marriage market}
 Throughout this paper, I consider a frictionless\footnote{For state of the art applications of the search and matching models to the marriage market, see \textcite{GousseJacquemetRobin2017} and \textcite{ColesFrancesconi2019}.}, bipartite, bilateral one-to-one marriage market. It is populated by men and women, indexed by $i\in \mathcal{I}$ and $j\in \mathcal{J}$, respectively, who may decide to form heterosexual pairs\footnote{I choose to ignore same-sex couples, as they account for only $1\%$ of couple households in the US in 2010. See \textcite{CiscatoGalichonGousse2020} for an empirical analysis of the marriage patterns of same-sex couples.}. A matching is a dummy variable $\mu_{ij}$ that is equal to one if man $i$ and woman $j$ marry, and zero otherwise.

 When staying single, agents receive payoffs $\mathcal{U}_{i0}$ and $\mathcal{V}_{0j}$ for men and women, respectively. Whenever a man $i$ and a woman $j$ meet on the marriage market, they bargain over utility outcomes $(u,v)$ that lie in a bargaining set $\mathcal{F}_{ij}$.

\subsection{Proper bargaining sets}
\begin{figure}[h]
  \caption{A typical proper bargaining set $\mathcal{F}_{ij}$}\label{fig:bargainingsetF}
\centering
  \includegraphics[width=0.5\textwidth, trim=1cm 1cm 1cm 1cm,clip]{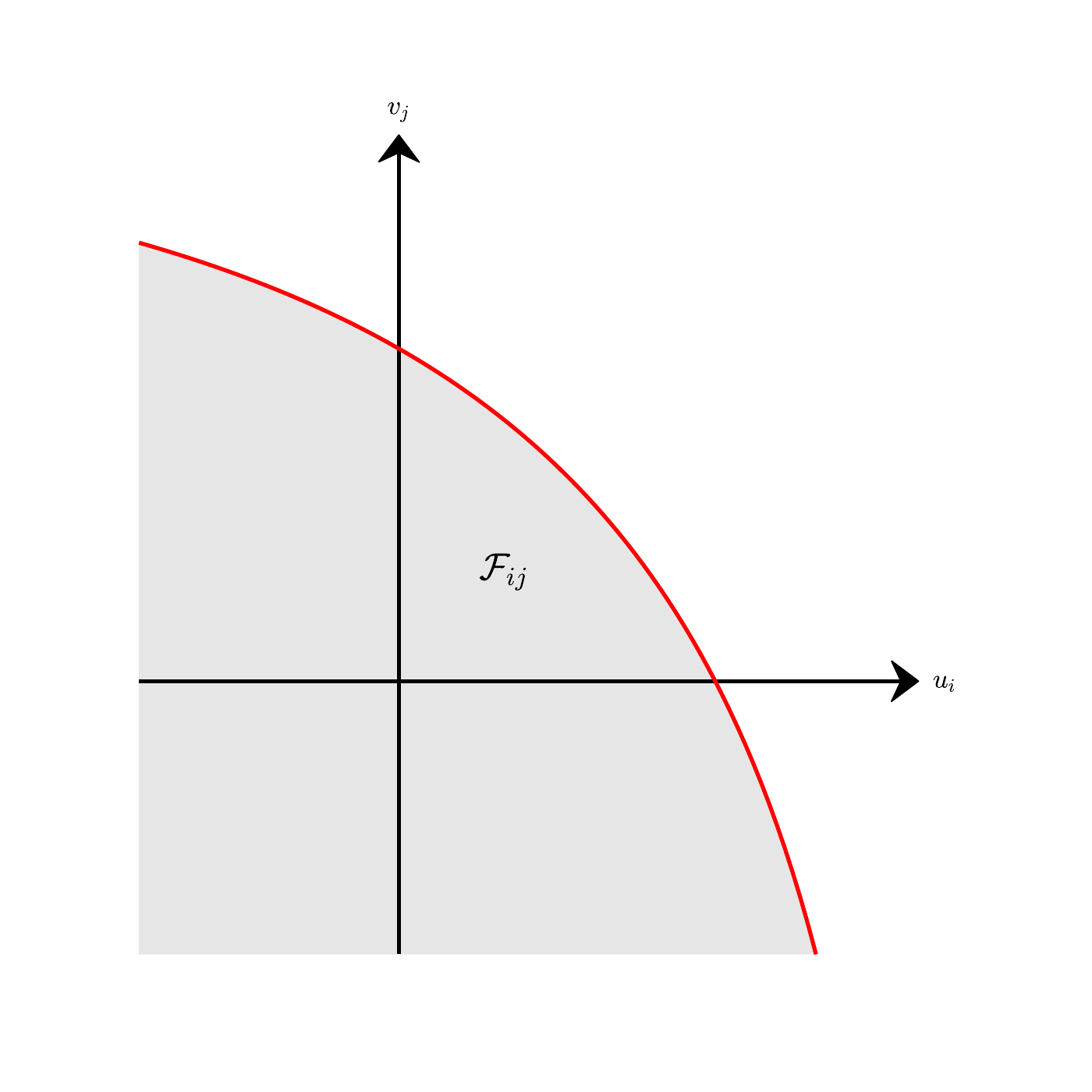}
\end{figure}

\newcommand{\setG}{\textit{utility possibility set}}
\newcommand{\setH}{\textit{strict utility possibility set}}
The starting point in GKW are the bargaining sets, which serve as primitives of the model. It is assumed that these sets meet a certain number of requirements, in which case they are called \textit{proper bargaining sets} (see definition 1 in GKW). For convenience, the definition of a proper bargaining set is reproduced below, while figure \ref{fig:bargainingsetF} shows a typical proper bargaining set.

\begin{definition}[Galichon, Kominers and Weber, 2018]
A bargaining set $\mathcal{F}_{ij}$ is a
\emph{proper bargaining set} if the three following conditions are met: (i)
$\mathcal{F}_{ij}$ is closed and nonempty (ii) $\mathcal{F}_{ij}$ is \emph{lower
comprehensive}: if $\left(  u,v\right)  \in\mathcal{F}_{ij}$, then $\left(
u^{\prime},v^{\prime}\right)  \in\mathcal{F}_{ij}$ provided $u^{\prime}\leq u$
and $v^{\prime}\leq v$. (iii) $\mathcal{F}_{ij}$ is \emph{bounded above}:
Assume $u_{n}\rightarrow+\infty$ and $v_{n}$ bounded below then for $N$ large
enough $\left(  u_{n},v_{n}\right)  $ does not belong in $\mathcal{F}$ for
$n\geq N$; similarly for $u_{n}$ bounded below and $v_{n}\rightarrow+\infty$.
\end{definition}

\subsection{Parametrizing the bargaining sets via distance functions}

As in most matching frameworks, we will write down stability conditions in order to define equilibrium on the marriage market. However, this is not as trivial in the ITU case as it is with perfectly transferable utility. To reach our goal, we will make use of a parametrization of the bargaining sets that we call \emph{distance functions}.

\begin{figure}
  \centering
  \includegraphics[width=0.5\textwidth, trim=1cm 1cm 1cm 1cm,clip]{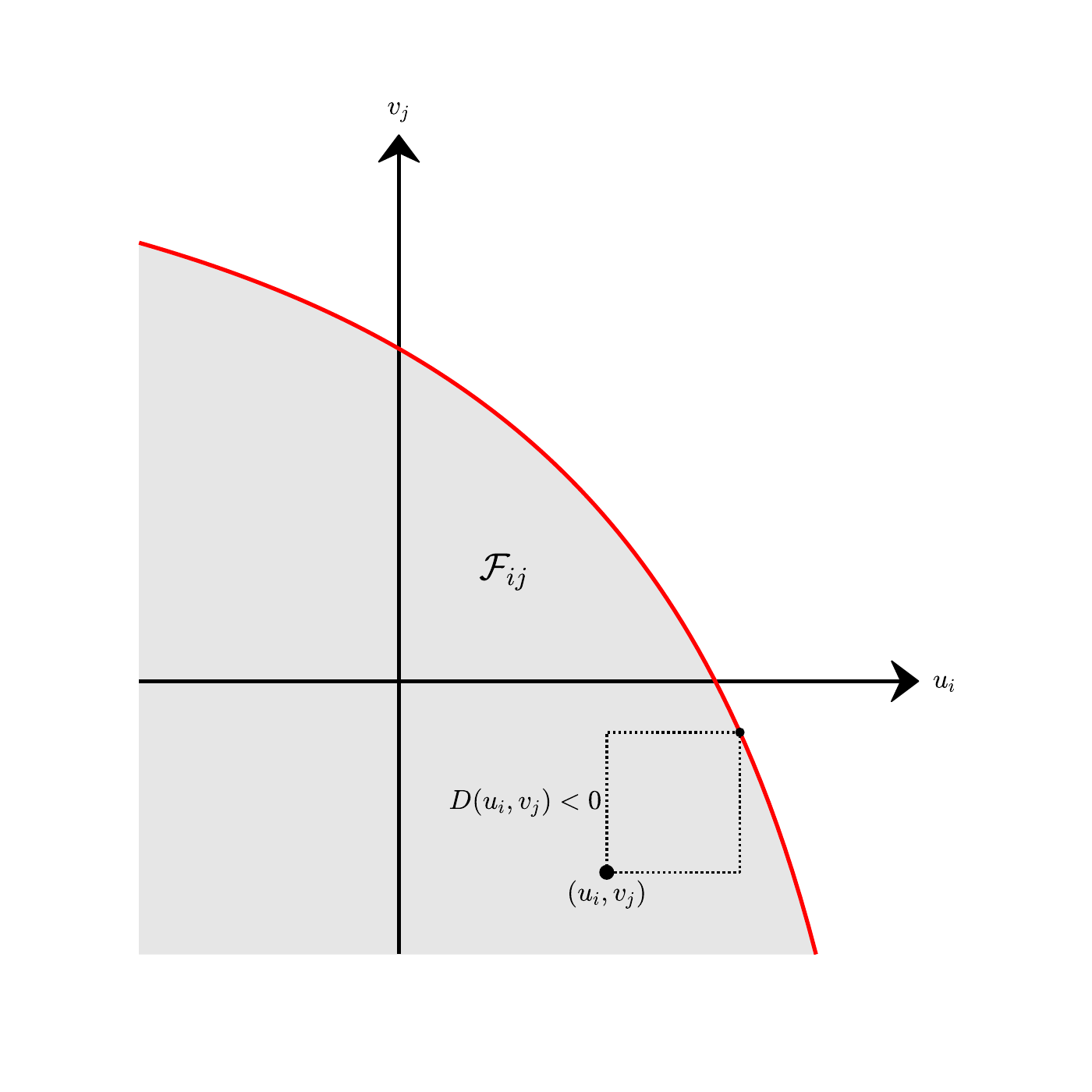}
  \caption{The distance function for a feasible point $(u_i,v_j)$}\label{fig:distance}
\end{figure}

The \textit{distance function} $D$ is an analytical tool that describes whether a utility allocation $(u,v)$ is a boundary point of the bargaining set or not. The distance function is not only useful to characterize the equilibrium in a simpler way, but also turns out to be crucial for estimation purposes. For a given pair $(i,j)$ and a generic bargaining set $\mathcal{F}_{ij}$, the distance function is defined as
\[
D_{ij}\left(  u,v\right)  =\min\left\{  z\in\mathbb{R}:\left(
u-z,v-z\right)  \in\mathcal{F}_{ij}\right\}  .
\]
Note that, indeed, $D_{ij}(u,v)=0$ whenever the pair of utilities $(u,v)$ belongs to the
boundary of $\mathcal{F}_{ij}$, $D_{ij}(u,v)\leq0$ if the pair $(u,v)\,$\ is feasible, and
$D_{ij}(u,v)>0$ otherwise. This is illustrated in figure \ref{fig:distance} for a feasible point $(u_i,v_j)$.

Stability conditions require that (i) for all men $i$ and all women $j$, $u_i\geq \mathcal{U}_{i0}$ and $v_j\geq\mathcal{V}_{0j}$ (these are the usual participation constraints) and (ii) that for all $i,j$ pairs, the equilibrium payoffs $u_i$ and $v_j$ satisfy
\begin{equation}\label{eq:stability}
  D_{ij}(u_i,v_j)\geq0
\end{equation}
Note that condition (\ref{eq:stability}) holds with equality whenever man $i$ and woman $j$ are married in equilibrium. The intuition behind the stability conditions\footnote{Note that in a model with transferable utility where partners share a surplus $\Phi_{ij}$, the distance function is $D_{ij}(u,v) = (u_i+v_j-\Phi_{ij})/2$, so that we recover the usual stability condition $u_i+v_j\geq \Phi_{ij}$.} is as follows. Assume that (\ref{eq:stability}) does not hold for some $(i,j)$ pair, that is, $D_{ij}(u_i,v_j)<0$ (such a point is represented in figure \ref{fig:distance}). But then, whenever man $i$ and woman $j$ meet, they can choose a feasible utility allocation so that both of them are strictly better off (any point in the square area in figure \ref{fig:distance} satisfies this). Hence, man $i$ and woman $j$ form a blocking pair.

\subsection{Introducing heterogeneity in tastes}
Starting from section \ref{sec:estimation} in the present paper, I will make use of the same empirical framework as in \textcite{GalichonKominersWeber2019}. Namely, I will introduce \emph{types} and heterogeneity in tastes.

Types are groups of men and women who share similar observable characteristics.  The sets of types are denoted $\mathcal{X}$ for men and $\mathcal{Y}$ for women, while $\mathcal{X}_{0}=\mathcal{X}\cup \mathcal{\{}0%
\mathcal{\}}$ and $\mathcal{Y}_{0}=\mathcal{Y}\cup \mathcal{\{}0\mathcal{\}}$
introduce singlehood as an option (denoted $\{0\}$). I let $n_{x}$ be the mass of men of type $x$, and $m_{y}$ be the mass of women
of type $y.$ A matching is now a vector $(\mu _{xy})_{x\in \mathcal{X},y\in \mathcal{Y}}$ that gives the mass of matches between men of type $x$ and women of type $%
y$. The vectors $(\mu _{x0})_{x\in \mathcal{X}}$ and $(\mu _{0y})_{y\in \mathcal{%
Y}}$ denote the mass of single men of type $x$ and single women of type $y$,
respectively. Of course, a feasible matching must satisfy the following scarcity constraints

\begin{equation}\label{eq:scarcity}
\begin{split}
  \mu_{x0} + \sum_y \mu_{xy} &= n_x \\
  \mu_{0y} + \sum_x \mu_{xy} &= m_y
\end{split}
\end{equation}

Heterogeneity in tastes is introduced as follows. As in GKW, I assume that the payoffs received by man $i$ and woman $j$ are the sum of two components: (i) a systematic part, denoted $U_i$ and $V_j$, respectively, that lie within the bargaining set $\mathcal{F}_{ij}$ and over which the potential partners bargain, and (ii) an idiosyncratic part, denoted $\epsilon_{iy_j}$ and $\epsilon_{x_ij}$, respectively. In addition, it is assumed that the bargaining set only depends on the observable types, that is, $\mathcal{F}_{ij} = \mathcal{F}_{x_iy_j}$\footnote{Together, these assumptions are very similar to the ``separability'' assumption in the classical transferable utility framework with unobserved heterogeneity. See \textcite{GalichonKominersWeber2019} for a discussion.}.

One contribution of GKW is to show that we can actually focus on equilibrium where the systematic utilities, that I will denote $U_{x_i y_j}$ and $V_{x_i y_j}$, depend on the observable types of the partners. They show that doing so is in fact merely a restriction, as there always exists an individual equilibrium of this form. The advantage of focusing on such equilibria is that we can reduce the matching problem to a series of discrete choice problems, where each individual chooses the observable type of his potential partner.

To simplify further the analysis, I will assume that the taste shocks are i.i.d draws from an extreme value type I distribution. In this case, coined the ``ITU-logit model'', it is well known that the systematic utilities can be obtained from
the matching patterns using the log odds ratio formula (see \textcite{McFadden1974}, and \textcite{ChooSiow2006} for an application to the marriage market). Namely:
\begin{equation*}
U_{xy} = \log\frac{\mu_{xy}}{\mu_{x0}} \text{ and } V_{xy} = \log\frac{\mu_{xy}}{\mu_{0y}}
\end{equation*}
In this setting, characterizing equilibrium is very simple. Indeed, the triple $(\mu_{xy}, U_{xy}, V_{xy})$
is an aggregate equilibrium (see definition 5 in GKW) if and only if the systematic utilities given by $U_{xy} = \log\frac{\mu_{xy}}{\mu_{x0}}$ and  $V_{xy} = \log\frac{\mu_{xy}}{\mu_{0y}}$ satisfy
\begin{equation}\label{eq:aggreq}
  D_{xy}(\log\frac{\mu_{xy}}{\mu_{x0}}, \log\frac{\mu_{xy}}{\mu_{0y}}) = 0
\end{equation}
and the matching satisfies the scarcity constraints (\ref{eq:scarcity}). Equation (\ref{eq:aggreq}) describes the usual feasibility/stability conditions, while the system of equations in (\ref{eq:scarcity}) ensures the the equilibrium matching is feasible, given the population supplies $(n_x,m_y)$.

Combining these conditions, we get that in the ITU-logit framework, equilibrium is fully characterized by the system of equations
\begin{equation}\label{eq:estim2}
\begin{split}
  \mu_{x0} + \sum_y \exp(-D_{xy}(-\log \mu_{x0},-\log\mu_{0y})) &= n_x \\
  \mu_{0y} + \sum_x \exp(-D_{xy}(-\log \mu_{x0},-\log\mu_{0y})) &= m_y
\end{split}
\end{equation}
In section \ref{sec:estimation}, I will provide a new estimation method based on the above characterization of equilibrium.

Before going into section \ref{sec:bargaining}, I shall point out that while I stick to the vocabulary of marriage markets throughout the paper, my results apply to other matching markets as well, e.g. between workers and firms.

\section{Proper collective models}\label{sec:bargaining}

Two steps are required to link collective models and models of matching with imperfectly transferable utility. First, assuming that we have some collective model in mind, we must show that the bargaining sets it generates are proper, as required by the \textcite{GalichonKominersWeber2019} matching framework. Second, since in equilibrium partners select a utility allocation that belongs to the frontier of the bargaining set, we must show that such a point is Pareto efficient. If these two conditions are met, then the collective model is said to be \emph{proper}.

\subsection{The primitives}
The bargaining sets are, in GKW, the primitives of the model. In the present framework, however, the bargaining sets are not given. Instead, I assume that a specific collective model is the primitive. By that, I mean that men and women have preferences over the observable characteristics of their partner (non-economic marital gains) as well as over goods (private and possibly public goods as well) that they wish to consume (economic gains to marriage). Whenever man $i$ meet woman $j$, they bargain and choose an allocation of goods $\omega$ from a feasible set $\Omega_{ij}$, which depends on the characteristics of the partners (for example, wages and income) and technology available to them. A typical element of $\Omega_{ij}$ is an allocation $\omega=(q^a, q^b, Q)$, where $q^a$ and $q^b$ are vectors of private goods, consumed by the man and the woman respectively, and $Q$ a vector of public goods.
Man $i$ and woman $j$ receive utility $U_{ij}(q^a,Q)$ and $V_{ij}(q^b,Q)$, respectively. With some slight abuse of notation, I will drop the subscript $ij$ in the remainder of the section, since we will be focusing on only one $ij$ pair.

\begin{remark}[Preferences]
 Note that preferences are egotistic. Note also that I am agnostic about where the goods are coming from: some could be bought on the market, some others could be produced within the household. The model can be extended to accommodate for the presence of other household members (children, for example). However, I assume that there are only two decision makers in the household, namely the husband and the wife (collective models can have more than two decision makers, see e.g. \textcite{DauphinLahgaFortinEtAl2011}).
\end{remark}

Thus, in this framework, the preferences of men and women (represented by the utility functions $U$ and $V$) and the set of feasible allocations of goods, $\Omega$, are primitives (which I summarize by saying that the collective model is the primitive). From them, we can derive the bargaining set as in GKW. As we shall see below, we can show that the resulting bargaining set is proper under mild assumptions. These assumptions are as follows:

\begin{assumption}[Feasible Allocations]\label{ass:O}
The set of feasible good allocations, denoted $\Omega$, is compact and convex.
\end{assumption}

\begin{assumption}[Utility Functions]\label{ass:UV}
Preferences are represented by utility functions $U$ and $V$ that are upper semi continuous and bounded above on $\Omega$ and strictly increasing.
\end{assumption}

Assumptions \ref{ass:O} and \ref{ass:UV} are fairly standard in microeconomic theory. With the usual linear inequality budget and time constraints, the set $\Omega$ is indeed convex and compact. The assumption is still holding when some goods are home-produced and the production function is assumed to be concave. As we shall see, upper semi continuity is enough to prove closedness of the bargaining set. I also make the assumption that agents consume only ``goods'' (as opposed to ``bads'') so that the utility is increasing in consumption. There are no negative externalities of consumption from any of the partners in this framework.

\subsection{From the primitives to the bargaining sets under free disposal}

As a first step, I make an assumption that will render trivial the construction of the bargaining sets and showing that they are proper, but will be relaxed later on. I do so to show that my approach (and the results that follow) is very general if one is willing to make this assumption. I assume that partners can engage in utility burning, i.e there is free disposal of utility. This assumption is discussed in the matching literature, see e.g. \textcite{GalichonHsieh2017}.

\begin{assumption}[Free disposal]\label{ass:free}
There is free disposal of utility. That is, if $u^\prime \leq u$, $v^\prime\leq v$ and if $(u, v)$ is feasible, then $(u^\prime, v^\prime)$ is feasible.
\end{assumption}

\begin{figure}[h]
  \caption{The bargaining set $\mathcal{G}$}\label{fig:bargainingsets}
\centering
  \includegraphics[width=0.5\textwidth, trim=1cm 1cm 1cm 1cm,clip]{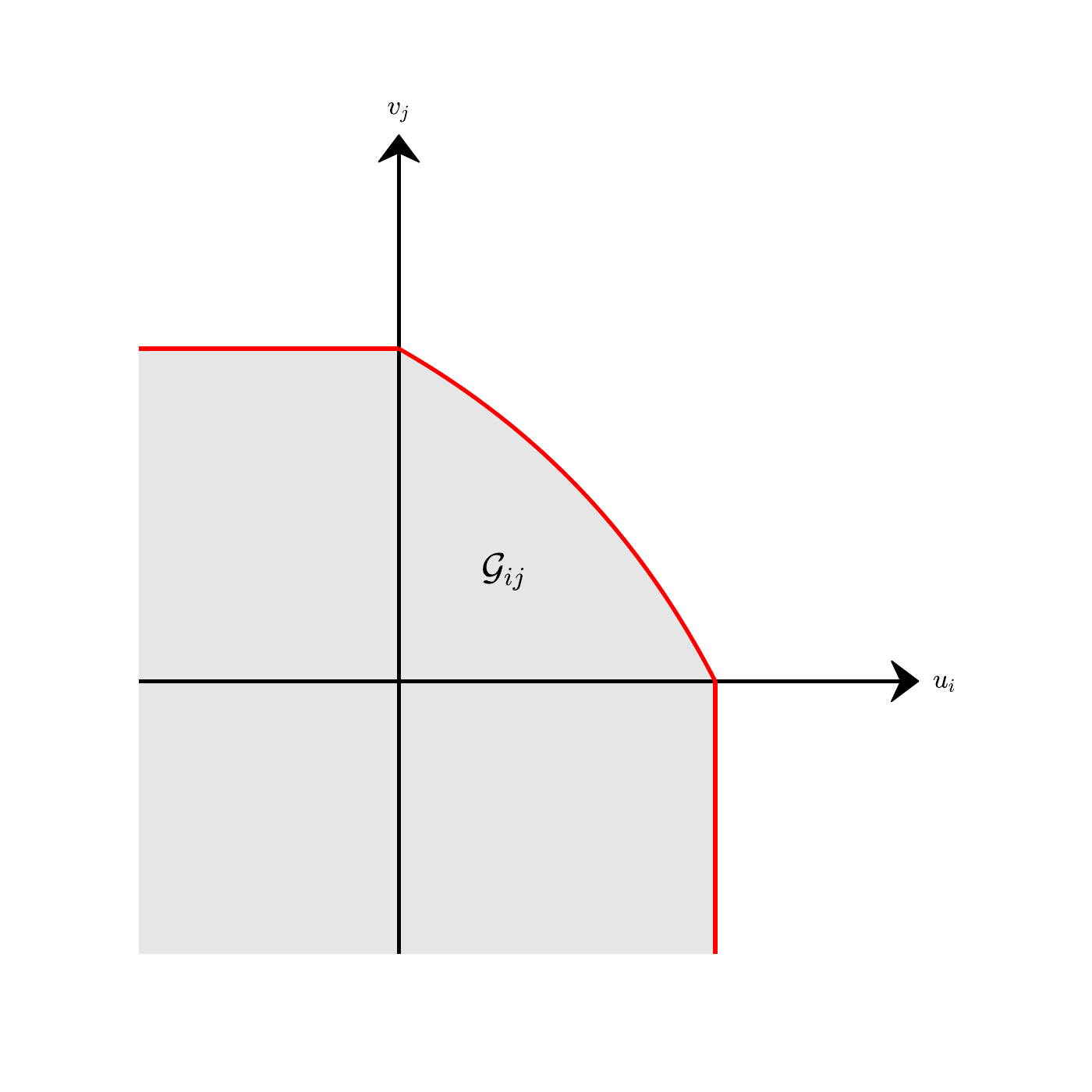}
\end{figure}

Assumption \ref{ass:free} means that we can construct a bargaining set, denoted $\mathcal{G}$, associated with $(\Omega,U,V)$ as the {\setG}
  \[
  \mathcal{G}=\{(u,v):\exists (q^a,q^b,Q)\in\Omega:u\leq
U(q^a,Q),v\leq V(q^b,Q)\}
\]
By definition then, this set satisfies lower-comprehensiveness. A graphical example of the set $\mathcal{G}$ is given in figure \ref{fig:bargainingsets}. We need only to show that this bargaining set satisfies the other requirements to be proper.

\begin{proposition}
\label{prop:Fproper} Under assumptions \ref{ass:O}, \ref{ass:UV} and \ref{ass:free}, the bargaining set $\mathcal{G}$ associated with $(\Omega,U,V)$ is a proper bargaining set
\end{proposition}

We have just shown that under mild conditions, we can actually construct the bargaining sets primitives in GKW as the \setG s associated to the underlying collective model, and these sets are proper. We also know that in the ITU setting, equilibrium utilities (payoffs) belong to the boundary of the bargaining set. What can we say about these points in terms of Pareto efficiency?

\begin{proposition}
\label{prop:weakPE} Whenever the set $\mathcal{G}$ is a proper bargaining set, a boundary point of $\mathcal{G}$ is weakly Pareto efficient.
\end{proposition}

\subsection{Relaxing free disposal}

The assumptions made in the previous section allow us to cover a great variety of models. However, if we are to focus on a typical collective model as primitive, we must relax the free disposal assumption. I will also assume that we choose a strictly concave cardinal representation of utilities. In this section, we will have to proceed differently to construct the bargaining sets, but under a few additional assumptions it can be shown that they are proper as well.

\begin{figure}[h]
\caption{The bargaining set $\mathcal{H}$}\label{fig:bargainingsetH}
\centering
\begin{subfigure}[b]{0.49\textwidth}
  \includegraphics[width=\textwidth, trim=1cm 1cm 1cm 1cm,clip]{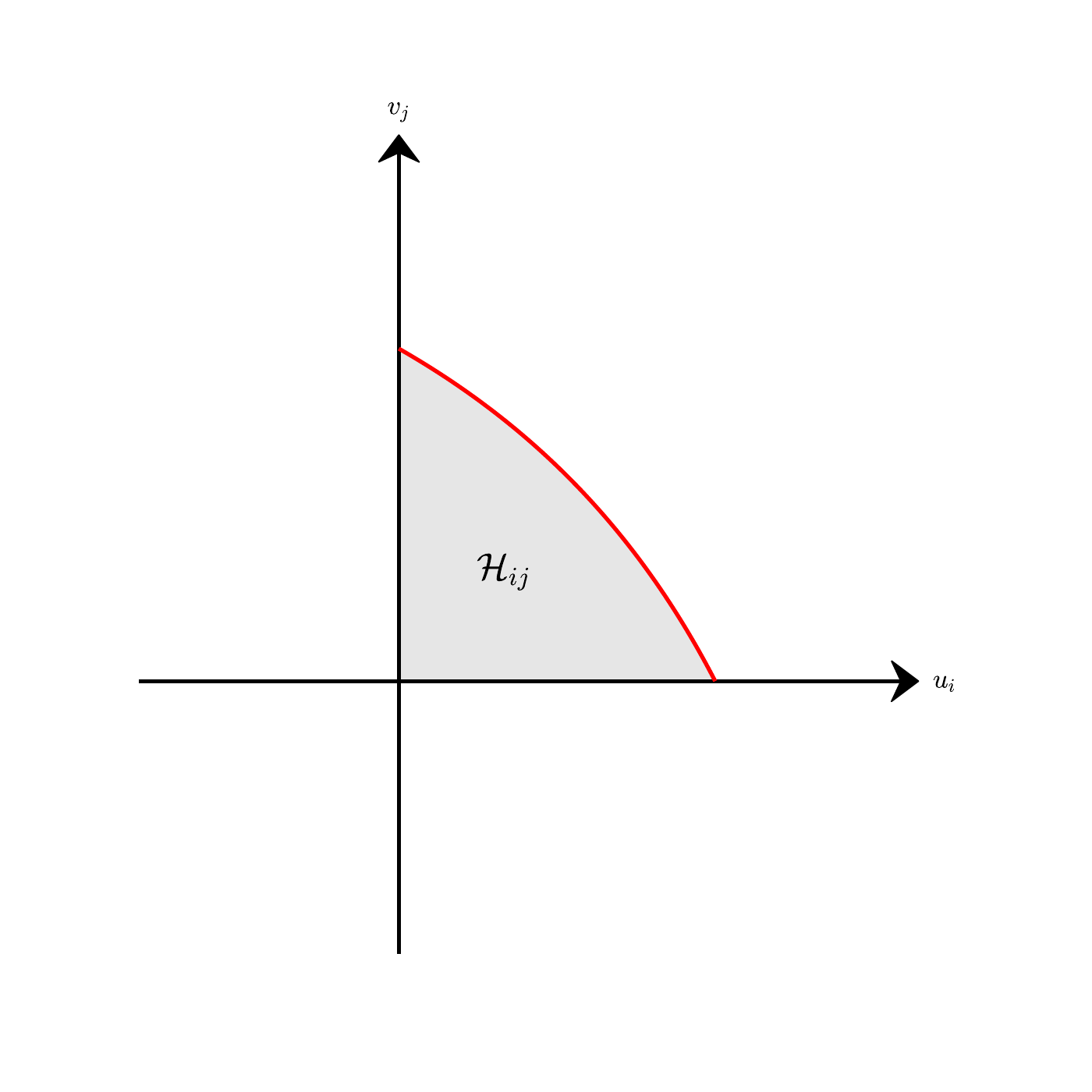}
  \caption{Improper}
  \end{subfigure}
  \begin{subfigure}[b]{0.49\textwidth}
  \includegraphics[width=\textwidth, trim=1cm 1cm 1cm 1cm,clip]{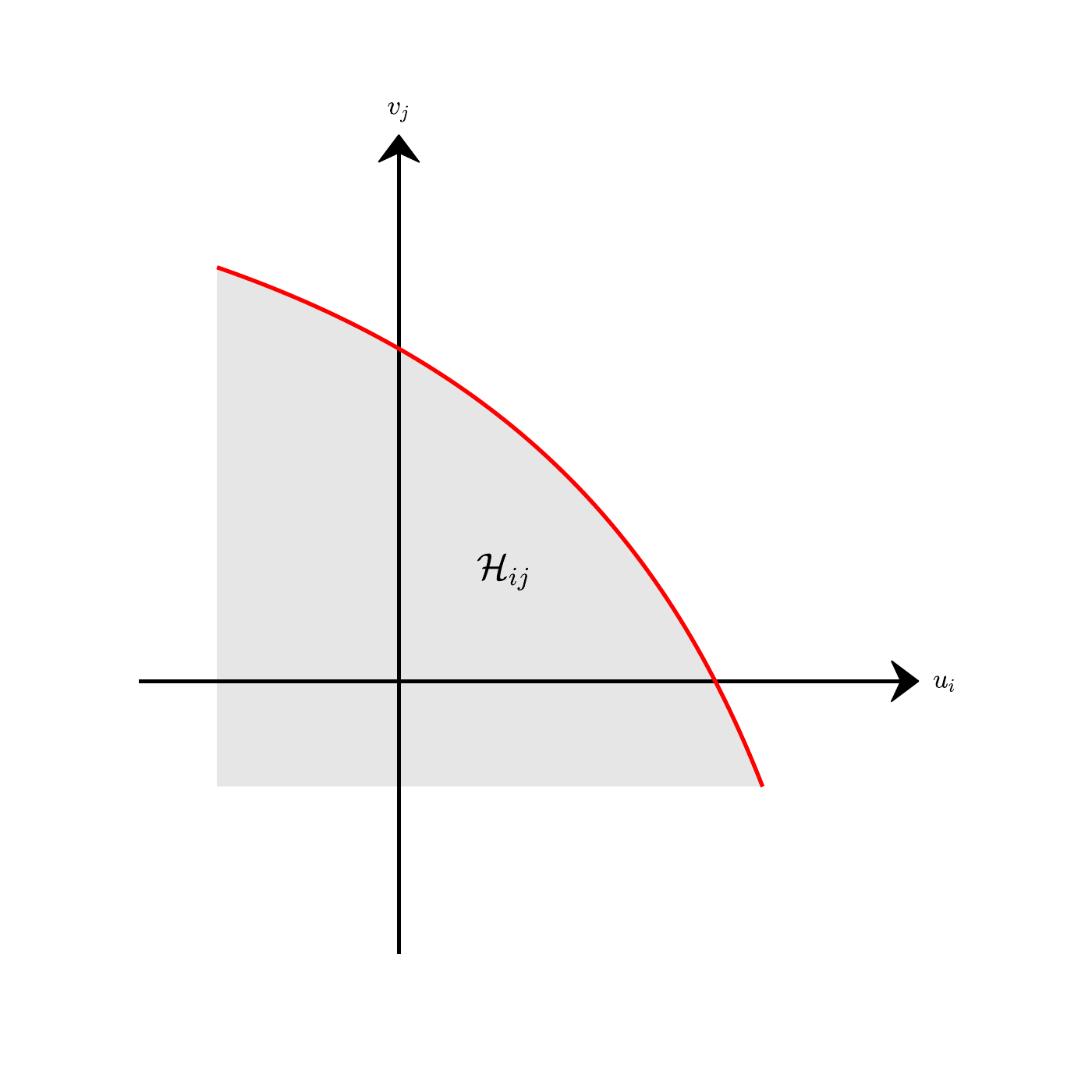}
  \caption{Proper}
  \end{subfigure}

\end{figure}

First, suppose assumptions \ref{ass:O} and \ref{ass:UV} hold, and construct the following bargaining set $\mathcal{H}$ associated with $(\Omega,U,V)$ as the {\setH} $\mathcal{H}$
\[
  \mathcal{H} = \{(u,v):\exists (q^a,q^b,Q)\in\Omega:u=
U(q^a,Q),v= V(q^b,Q)\}
\]
Figure \ref{fig:bargainingsetH} shows two examples of a bargaining set $\mathcal{H}$, one being proper, the other one is not.

Showing that the bargaining set $\mathcal{H}$ is proper is, unfortunately, impossible without additional assumptions. And, once again, if we want to connect the ITU framework to collective models, we need to provide the following two results: (i) if $\mathcal{H}$ is a proper bargaining set, then a boundary point is Pareto efficient, and (ii) $\mathcal{H}$ is indeed a proper bargaining set.

Let us start by showing (ii), assuming that the set $\mathcal{H}$ is a proper bargaining set. We will make the following two assumptions:
\begin{assumption}[Transferability]
\label{ass:transfers}If $(q^a,q^b,Q)\in\Omega$ and $q^a_{k}>0$ for some
private good $k$, then for any $q^{a^\prime}$ such that $q_{-k}^{a^\prime}=q^a_{-k}%
$, $q_{k}^{a^\prime}<q^a_{k}$ and $(q^{a^\prime},q^b,Q)\in\Omega$, there is a
$q^{b\prime}$ and some private good $l$ such that $q_{-l}^{b\prime}=q^b_{-l}$,
$q_{l}^{b\prime}>q^b_{l}$ and $(q^{a^\prime},q^{b\prime},Q)\in\Omega$.
\end{assumption}

\begin{assumption}[Concave cardinal representation]
\label{ass:concave}
  $U$ and $V$ are strictly quasi-concave
\end{assumption}

Assumption \ref{ass:transfers} states that for any feasible allocation in which the man (or the woman) consumes a strictly positive amount of some private good, we can always find a feasible allocation that is identical except for the fact that we slightly decreased the consumption of the man for that private good and slightly increased the consumption of the woman for some private good. All in all, assumption \ref{ass:transfers} states that there is always some ways (even if imperfect) of transferring utility from one partner to the other. Assumption \ref{ass:concave} is again very standard in the collective model literature. This will ensure that the bargaining sets are convex, which was not required in the previous section. We can now get the following result:

\begin{proposition}\label{prop:strongPE}
Whenever the {\setH} $\mathcal{H}$ is a proper bargaining set, and under assumption
\ref{ass:O}, \ref{ass:UV}, \ref{ass:transfers}, and \ref{ass:concave}, a
boundary point $(u,v)$ of $\mathcal{H}$ is strongly Pareto efficient.
\end{proposition}

Using the {\setH} $\mathcal{H}$ as our bargaining set is attractive because under the conditions mentioned above, the ITU framework imposes that in equilibrium partners choose a Pareto efficient outcome. Hence, the remaining piece of the puzzle is showing that we can indeed use $\mathcal{H}$ as a bargaining set by proving that it is proper.
\begin{assumption}[Vital goods]
\label{ass:limit}
There exists two private goods, $q^a_{1}$ and $q^b_{1}$ such that

(i) $\lim_{q^a_{1}\rightarrow0^{+}}U(q^a,Q)=-\infty$ and $\lim_{q^b_{1}%
\rightarrow0^{+}}V(q^b,Q) = -\infty$

(ii) for any $(q^a,q^b,Q)\in\Omega$, $((q_{1}^{a^\prime},q^a_{-1}),(q_{1}%
^{b\prime},q^b_{-1}),Q)\in\Omega$ whenever $q_{1}^{a^\prime}\leq q^a_{1}$ and
$q_{1}^{b\prime}\leq q^b_{1}$
\end{assumption}

Assumption \ref{ass:limit} states that both men and women consume a vital private good, in the sense that if the quantity consumed tends to zero, their utility tends to minus infinity. It also states that if the household decreases the quantity of these goods, he can still at least buy the same amount of the other goods. In effect, this assumption plays the same role than the free-disposal restriction: if the couple agrees on an utility allocation $(u,v)$, it is always possible to reach an allocation $(u^\prime,v^\prime):u^\prime\leq u,v^\prime\leq v$ by decreasing the amount of vital goods consumed. In practice, one can think of food or water filling this role\footnote{Although food preparation may be a public good, the consumption of food is private.}. We now reach our final result:

\begin{proposition}\label{prop:properfinal}
Under Assumption \ref{ass:O}, \ref{ass:UV} and \ref{ass:limit}, the {\setH} $\mathcal{H}$ associated with $(\Omega,U,V)$ is a
proper bargaining set
\end{proposition}

We can now combine all the results in the following corollary:
\begin{corollary}\label{prop:properpareto}
Under Assumption \ref{ass:O}, \ref{ass:UV}, \ref{ass:transfers}, \ref{ass:concave} and \ref{ass:limit}, the bargaining set $\mathcal{H}$ is a proper bargaining set and any point on the boundary of $\mathcal{H}$ is strongly Pareto efficient.
\end{corollary}

The proof of this corollary follows from putting together the previous results. For simplicity, I refer to collective models satisfying the assumptions invoked in corollary \ref{prop:properpareto} as \emph{proper collective models}. A formal definition is given below.
\begin{definition}\label{def:propercm}
  A collective model that satisfies assumptions \ref{ass:O}, \ref{ass:UV}, \ref{ass:transfers}, \ref{ass:concave} and \ref{ass:limit} is a \textit{proper collective model}. Indeed, the resulting bargaining set $\mathcal{H}$ is proper and convex and any point on the frontier is Pareto-efficient.
\end{definition}

\section{A computation method}\label{sec:compmethod}

From now on, I am assuming that we are equipped with a proper collective model. We have also chosen a convenient functional form for the utility functions of men and women. The (preference) parameters that enter the utility functions are gathered in a vector denoted $\theta$\footnote{For example, we may assume that $U(q^a,Q) = \theta_0 + \theta_1\log q^a + \theta_2 \log Q$, so that $\theta=(\theta_0,\theta_1,\theta_2)$.}. In this section, I will discuss how to compute the distance function in the bargaining sets associated to such model ($\theta$ being given). In section \ref{sec:estimation}, I will discuss how to estimate $\theta$.

\subsection{Computational issues behind distance functions}
Unfortunately, distance function are not, in general, known in closed-form. In fact, as soon as the underlying collective model features multiple budget and time constraints, non-trivial preferences for public goods, household production or has possibly corner solutions, then it is not possible to obtain an analytical formula for the distance function. This is best illustrated by the following example.

\begin{example}\label{ex:dist}
  Consider a case in which utility depends on the consumption of a private good $q$ and a public good $Q$. All prices are set to one. The systematic utilities are as follows
  \begin{align*}
    U_{ij}(q^a_{ij},Q_{ij}) = \delta^a_{ij} + \log q^a_{ij} + A \log Q_{ij} \\
    V_{ij}(q^b_{ij},Q_{ij}) = \delta^b_{ij} + \log q^b_{ij} + B \log Q_{ij}
  \end{align*}
  where I assume $A\neq B$. The terms $\delta^a_{ij}$ and $\delta^b_{ij}$ are non-economic gains to marriage. Normalizing prices to one and denoting $\Phi_{ij}$ the total income of pair $(i,j)$, the budget constraint is $q^a_{ij} + q^b_{ij} + Q_{ij} \leq \Phi_{ij}$. According to section \ref{sec:bargaining}, the bargaining set $\mathcal{H}_{ij}$ associated with this model is a proper bargaining set. To compute the distance function, we make use of the first order conditions (FOCs) in the corresponding collective model. We get the standard Bowen-Lindhal-Samuelson (BLS) condition
  \[
  A q^a_{ij} + B q^b_{ij} = Q_{ij}
  \]
  Say we want to compute the distance at point $(u,v)$, that is, we want to find $z$ such that $(u-z,v-z)$ belongs to the frontier. It is easily seen that the BLS condition defines $Q_{ij}$ as an implicit function of $(u-z,v-z)$ with
  \begin{equation}
\begin{split}
  A \exp(u-z-\delta^a_{ij}-A\log Q_{ij}(u-z,v-z)) + \\
  B \exp(v-z-\delta^b_{ij}-B\log Q_{ij}(u-z,v-z)) = Q_{ij}(u-z,v-z)
\end{split}
\end{equation}
  Unless $A=B$, there is no closed-form formula for the optimal $Q_{ij}$ and therefore no closed-form for the distance function, which, in this case is

\begin{equation*}
\begin{split}
  \min_{z\in\mathbb{R}} \{z: & \exp(u-z-\delta^a_{ij}-A\log Q_{ij}(u-z,v-z)) \\
                                & + \exp(v-z-\delta^b_{ij}-B\log Q_{ij}(u-z,v-z)) + Q_{ij}(u-z,v-z) \leq \Phi_{ij}\}
\end{split}
\end{equation*}

This example brings two remarks to mind. First, there is little computational gains in writing down the collective model and making use of the FOC (such as the Bowen-Lindhal-Samuelson condition). Indeed, we still need to minimize over $z$ and for each value of $z$ we must solve for the optimal $Q$ using numerical methods. Second, this example is still relatively simple because we only need to make use of one constraint (the budget constraint) to compute the distance function. We do not need worry about non negativity constraints since we work with logs, and there are no time constraints, corner solutions or public good production here. As we shall see, there are no easy way of solving more complex models.
\end{example}

\subsection{Method}
In the following, I introduce a method to compute distance functions when bargaining sets are constructed from proper collective models. In appendix \ref{app:comp}, I show that a similar method can be used whenever a bargaining set is proper (and so the method is not restricted to those bargaining sets generated by proper collective models). As we shall see below, computation relies on the very definition of the distance function but has interesting implications. While I keep things general, I will assume that the set of feasible allocations $\Omega_{ij}$ is fully characterized by $R$ feasibility constraints
\begin{equation}
g_{ij}^r(q^a_{ij},q^b_{ij},Q_{ij}) \leq0,\text{ }r\in\{1,...,R\}
\end{equation}
where $\{g_{ij}^r\}_{r\in\{1,...,R\}}$ are convex functions\footnote{Typically, the $\{g_{ij}^r\}_{r\in\{1,...,R\}}$ functions will be the usual linear budget and time constraints, but could also contain the production function of home-produced public goods.}.

\begin{problem}\label{eq:P1}
  Suppose that $\mathcal{H}_{ij}$ is a proper bargaining set associated with some proper collective model, with preferences and set of feasible goods $(U_{ij}^\theta, V_{ij}^\theta, \Omega_{ij})$. Then we can solve
 \begin{align}
&  \min_{z_{ij},q^a_{ij},q^b_{ij},Q_{ij}}z_{ij}\\
\begin{split}\label{eq:CompMethodii}
s.t\text{ } u-z_{ij}  &  =U_{ij}^{\theta}(q^a_{ij},Q_{ij})\\
            v-z_{ij}  &  =V_{ij}^{\theta}(q^b_{ij},Q_{ij})\\
            g_{ij}^r(q^a_{ij},q^b_{ij},Q_{ij})  &  \leq0,\text{ }r\in\{1,...,R\}
\end{split}
\end{align}
when $D^{\theta}_{ij}(u,v) = z_{ij}^*$, solution to the above program.
\end{problem}

In practice, it is very fast to solve such optimization programs, and given the
solution $z_{ij}^{\ast}$ to this problem, we have $D_{ij}(u,v)=z_{ij}^{\ast}$. To ease
computations, note that it is straightforward to provide the analytic gradient
of the objective function to the solver, and that, in general, the analytic expression of
the jacobian of the system of constraints can be obtained in analytic form as well.

\begin{remark}[Simplifying problem \ref{eq:P1}]
  The above technique is fairly general, but of course, it can be combined with the approach laid out in example \ref{ex:dist} to reduce the number of variables over which the minimization is performed. For instance, we saw in example \ref{ex:dist} that $Q_{ij}$ can be immediately deduced from $q^a_{ij}$ and $q^b_{ij}$ making use of the Bowen-Lindhal-Samuelson condition. Therefore, we can avoid minimizing over $Q_{ij}$ in problem \ref{eq:P1} above.
  In addition, solving problem \ref{eq:P1} might not be necessary when some goods are discrete, in which case one can apply lemma 2 from GKW. Lemma 2 in that paper states that if the proper bargaining set is the union (intersection) of basic proper bargaining sets, then the distance function is the minimum (maximum) over the distance functions associated to each of these basic sets. It may well be the case that these distance functions are known in closed-form when conditioning on the discrete good, in which case problem \ref{eq:P1} is not necessary. Therefore, lemma 2 is particularly useful for applications in which, for example, time is discrete or fertility is modelled.
\end{remark}

\begin{remark}[Parallel computing]
  Solving problem \ref{eq:P1} is simple, but the process has to be repeated many times and this is computationally intensive. On a small marriage market with $100$ men and $100$ women, we must compute the distance function 10,000 times. Fortunately, the process of computing the distance function can easily be parallelized. In addition, problem \ref{eq:P1} is solvable for multiple $(i,j)$ pairs at once. Let us denote $\mathcal{I}\mathcal{J}\equiv \{1,...,|\mathcal{I}||\mathcal{J}|\}$ the set indexing all possible $ij$ pairs and $\bar{\mathcal{I}\mathcal{J}}$ a subset of $\mathcal{I}\mathcal{J}$. To find the distance function for all $ij$ pairs in some set  $\bar{\mathcal{I}\mathcal{J}}$ at once, we need to minimize $\sum_{ij\in\bar{\mathcal{I}\mathcal{J}}} z_{ij}$ with respect to $\{z_{ij}, q^a_{ij}, q^b_{ij}, Q_{ij}\}_{ij\in\bar{\mathcal{I}\mathcal{J}}}$ subject to the constraints in (\ref{eq:CompMethodii}) written for each of the $ij\in\bar{\mathcal{I}\mathcal{J}}$ pairs. In practice, mixing these two approaches yields the best result. That is, we compute the distance function for subsets of $\mathcal{I}\mathcal{J}$ that form a partition of $\mathcal{I}\mathcal{J}$, when each subset is sent to a different worker.
\end{remark}

\subsection{Gradient of the distance function}

When the number of men and women on the market is large, it is computationally costly to obtain the derivative of the distance function for each pair $(i,j)$ using standard numerical methods. Fortunately, the following results shows how to obtain the derivative of the distance function $D^{\theta}_{ij}$ with respect to $u$, $v$ or $\theta$ at the cost of one evaluation of the distance function.

\begin{theorem}\label{thm:gradient}
Assume we are equipped with a proper collective model and associated proper bargaining set $\mathcal{H}_{ij}$. Introduce the Lagrange multipliers associated with the constraints in problem \ref{eq:P1} as $\lambda_{1}%
, \lambda_{2}$, and $\{\xi_{r}\}_{r\in\{1,...,R\}}$, respectively. The gradient with respect to $u,$ $v$ and $\theta$ of the distance function
$D_{ij}$ can be obtained directly when solving problem \ref{eq:P1}. Indeed, we have%
\begin{align*}
\partial_{u}D_{ij}^{\theta}(u,v)  &  =\lambda_{1}\\
\partial_{v}D_{ij}^{\theta}(u,v)  &  =\lambda_{2}\\
\partial_{\theta}D_{ij}^{\theta}(u,v)  &  =-\lambda_{1}\partial_{\theta
}U^{\theta}_{ij}(q^a_{ij},Q_{ij})-\lambda_{2}\partial_{\theta}V^{\theta}%
_{ij}(q^b_{ij},Q_{ij})
\end{align*}

\end{theorem}

\begin{remark}[Gradient of the decision variables] Solving problem \ref{eq:P1} also returns optimal values for $q^a_{ij}$, $q^b_{ij}$, and $Q_{ij}$. Can we compute the derivative of these values with respect to $u$, $v$ or $\theta$? While there is no general way of proceeding, it is usually possible. We can combine the first order conditions of our minimization problem and use the implicit function theorem to recover the derivatives. I will illustrate this point with an example. Suppose that the utility functions are given by
\begin{align*}
U_{ij}  &  =\delta^a_{ij} + a\log q^a_{ij}+A\log Q_{ij}\\
V_{ij}  &  =\delta^b_{ij} + b\log q^b_{ij}+B\log Q_{ij}
\end{align*}
and we are interested in computing the derivative of the optimal $q^a_{ij}$, $q^b_{ij}$, and $Q_{ij}$ with respect to $a$. Suppose that the only constraint is $q^a_{ij}+q^b_{ij}+Q_{ij} \leq \Phi_{ij}$ where $\Phi_{ij}$ is the income of pair $(i,j)$. We can see that in this model, the utility possibility set $\mathcal{H}_{ij}$ is a proper bargaining. It is easy to see that a solution to problem \ref{eq:P1} satisfies
\begin{align*}
q^a_{ij}+q^b_{ij}+Q_{ij} &  =\Phi_{ij}\\
u-D_{ij}^{\theta}(u,v) &  =a\log q^a_{ij}+A\log Q_{ij}\\
\frac{A}{a}q^a_{ij}+\frac{B}{b}q^b_{ij} &  =Q_{ij}
\end{align*}
where the first equality comes from the budget constraint (which is binding), the second is an equality constraint in problem \ref{eq:P1} and the third is obtained by combining the first order conditions from the underlying collective model. Therefore we have%
\[%
\begin{pmatrix}
\frac{a}{q^a_{ij}}-\frac{A}{Q_{ij}} & -\frac{A}{Q_{ij}}\\
\frac{A}{a}+1 & \frac{B}{b}+1
\end{pmatrix}%
\begin{pmatrix}
\partial_{a}q^a_{ij}\\
\partial_{a}q^b_{ij}%
\end{pmatrix}
=%
\begin{pmatrix}
-\partial_{a}D_{ij}^{\theta}(u,v)-\log q^a_{ij}\\
\frac{A}{a^{2}}q^a_{ij}%
\end{pmatrix}
\]
which we can rewrite $M\times\binom{\partial_{a}q^a_{ij}}{\partial_{a}q^b_{ij}}=d$,
so that%
\[
\binom{\partial_{a}q^a_{ij}}{\partial_{a}q^b_{ij}}=M^{-1}\times d
\]
and $\partial_a Q_{ij}$ follows easily.
\end{remark}

\subsection{Connection with collective models}

The computation methods introduced in the previous subsection highlights an interesting connection with collective models. It can be shown that the Lagrange multipliers associated with the constraint involving the utility functions of the partners in (\ref{eq:CompMethodii}) can be interpreted as Pareto weights. This relationship is proven in the following theorem:

\begin{theorem}\label{thm:collectivemodels}
Suppose we are equipped with a proper collective model and associated (proper) bargaining set $\mathcal{H}_{ij}$. Suppose that $\Omega_{ij}$ is characterized by a set of feasibility constraints $g_{ij}^r(q^a_{ij},q^b_{ij},Q_{ij})  \leq0,\text{ }r\in\{1,...,R\}$, where $\{g_{ij}^r\}$ are convex functions. Then

(i) given $u$, $v$ and $\theta$, the allocation $q_{ij}^{a\star},q_{ij}^{b\star},Q_{ij}^\star$ solution to problem \ref{eq:P1} is Pareto efficient.

(ii) in addition, the Pareto weights associated to the boundary point $(u-z^\star, v-z^\star)$ (reached via the allocation of goods $q_{ij}^{a\star},q_{ij}^{b\star},Q_{ij}^\star$) for the man and the woman are, respectively, the Lagrange multipliers $\lambda_1^\star$ and $\lambda_2^\star$. $\lambda_1^\star$ and $\lambda_2^\star$ also satisfy $\lambda_1^\star+\lambda_2^\star=1$
\end{theorem}

The proof of theorem \ref{thm:collectivemodels} makes use of the expression of the Lagrangian of the problem of computing the distance function. The main idea is to fix the irrelevant variables and recover the expression of the Lagrangian of a collective model problem. This result is important in theory as well as in practice, since it states that we can compute the Pareto weights from the Lagrange multipliers (which will be provided by any solver).

\begin{remark}[From bargaining sets to collective models]
Theorem \ref{thm:collectivemodels} illustrates that abstract concepts from GKW, e.g. distance functions, can be related to familiar objects of the collective framework. This is not surprising because whenever a collective model is proper, the Pareto weight associated to some boundary point of the Pareto set is in fact equal to (minus) the slope of the frontier at that particular point. And distance functions are nothing else than an alternative way of parametrizing the frontier. To see that, let us consider a proper collective model. The usual way of modelling efficiency is to solve
\[
\max_{(q^a,q^b,Q)\in\Omega} \mu/(1-\mu) U(q^a,Q) + V(q^b,Q)
\]
where $\mu/(1-\mu)$ is the Pareto weight. Or equivalently, we can solve
\[
\max_{(q^a,q^b,Q)\in\Omega} V(q^b,Q)
\]
subject to the constraint $U(q^a,Q) \geq u$ for some utility level $u$. In the latter formulation, it is clear that the Lagrange multiplier associated to the constraint $U(q^a,Q) \geq u$ is the Pareto weight $\mu/(1-\mu)$. I denote $\Psi(u)$ the value function in this second formulation of the optimization program, for some utility level $u$. Note that the equation of the Pareto frontier is therefore given by $v=\Psi(u)$. Hence, applying the envelop theorem, we see that minus the slope of the efficient frontier is in fact the Pareto weight, that is $-\partial_u \Psi(u) = \mu/(1-\mu)$. Since in this class of collective models, the Pareto frontier is strictly concave, decreasing and smooth, there is a one-to-one relationship between the Pareto weight and the position along the efficient frontier.

What theorem \ref{thm:collectivemodels} describes is the relationship between the Pareto weight and the implicit representation of the efficient frontier given by the distance function. Indeed we have, by definition,
\[
D(u,v)=0
\]
whenever $(u,v)$ belongs to the frontier, which again implicitly defines the mapping $\Psi:u\rightarrow v$ such that:
\[
D(u,\Psi(u))=0
\]
Differentiating, we get $\partial_u D + \partial_v D \partial_u \Psi = 0$, that is
\[
\mu/(1-\mu) \equiv -\partial_u \Psi(u) = \frac{\partial_u D}{\partial_v D}
\]
where we recover the result from theorem \ref{thm:collectivemodels}.

Aside from distance functions, the utility ``wedge'' introduced in GKW is another abstract object that proves to be connected to the notion of bargaining power. Let me first recall the definition of the wedge. For any proper bargaining set $\mathcal{F}$, GKW show that there exists two unique functions $\mathcal{U}$ and $\mathcal{V}$\footnote{These functions are $\mathcal{U}(w) = -D(0,-w)$ and $\mathcal{V}(w) = -D(w,0)$.} such that the set of frontier points $\{(u,v):D(u,v)=0\}$ is given by $\{(\mathcal{U}(w),\mathcal{V}(w)), w\in W\}$ for some open interval $W$ in $\mathbb{R}$. The parameter $w$, called ``wedge'', parametrizes the efficient frontier.
Suppose now that a bargaining set $\mathcal{F}$ is proper, and its frontier is smooth, strictly decreasing and concave. Making use of the explicit representation, we have
\begin{equation*}
D(\mathcal{U}\left( w\right) ,\mathcal{V}\left( w\right) )=0
\end{equation*}%
for all $w\in W$. Thus $\partial _{u}D\partial _{w}\mathcal{U}+\partial
_{v}D\partial _{w}\mathcal{V}=0$, hence%
\begin{eqnarray*}
\frac{\partial _{u}D}{\partial _{v}D} &=&-\frac{\partial _{w}\mathcal{V}}{%
\partial _{w}\mathcal{U}} \\
\mu/(1-\mu) &=&-\frac{\partial _{w}\mathcal{U-}1}{\partial _{w}\mathcal{U}} =\frac{1}{\partial _{w}\mathcal{U}}-1 \\
&=&-\frac{1}{\partial _{w}D(0,-w)}-1
\end{eqnarray*}
Therefore, in the context of proper collective models, the wedge and the Pareto weights are two equivalent ways of parametrizing the efficient frontier. It is true, however, that the approach using wedges is much more general because it can accommodate bargaining sets whose frontier features kinks and flat segments. In these cases, the approach via Pareto weights would fail because a point on the frontier can be associated with an infinity of Pareto weights (kinks) or a Pareto weight can be associated with an infinity of points on the frontier (flat segments). In contrast, the correspondence between the wedge and the location on the frontier is always one-to-one.
\end{remark}

\section{Estimation}\label{sec:estimation}

In this section, I present an new approach to estimate a proper collective model within the ITU-logit framework (see section 6 of GKW or section \ref{sec:preamble} of the present paper for a review), based on a mathematical program with equilibrium constraints. Note that section \ref{sec:bargaining} and appendix \ref{app:comp} imply that more general models could be estimated using the computation and estimation methods presented here, as long as they generate proper bargaining sets. For simplicity however, I focus on the case of proper collective models.

\subsection{Parameterization}
A key difference with the GKW setting is that I will use parameterizations of the systematic utilities $U_{xy}$ and $V_{xy}$ that are very common in the collective model literature. Specifically, I will choose a parametric proper collective model, following definition \ref{def:propercm}.

When a man of type $x$ meets with a woman of type $y$, they choose vectors of private consumption, denoted $q^a_{xy}$ and $q^b_{xy}$ respectively, as well as a vector of public consumption $Q_{xy}$. Naturally, the set of feasible allocations $(q^a_{xy},q^b_{xy},Q_{xy})$ must belong to some feasible set $\Omega_{xy}$ that satisfies assumption \ref{ass:O}. For a given allocation $(q^a_{xy},q^b_{xy},Q_{xy})$, the partners receive the following amounts of utility
\[
\mathcal{U}_{xy}^{\theta}(q^a_{xy},Q_{xy}) \text{ and } \mathcal{V}_{xy}^{\theta}(q^b_{xy},Q_{xy})
\]
where $\theta$ is a vector of preference parameters.

I shall also specify the outside options (singlehood) for a man of type $x$ and a woman of type $y$. I assume that they have access to a subset of the goods they can consume while married. They choose vectors of private and public goods, denoted $(q_x^s, Q_x^s)$ and $(q_y^s,Q_y^s)$ that must belong to some feasible sets $\Omega_{x0}$ and $\Omega_{0y}$, respectively. They receive utility
\[
\mathcal{U}_{x0}^{\theta}(q_x^s, Q_x^s) \text{ and } \mathcal{V}_{0y}^{\theta}(q_y^s,Q_y^s)
\]
Note that in general, it will be very straightforward to solve for the optimal consumption choices $(q_x^{s\star},Q_x^{s\star})$ and $(q_y^{s\star},Q_x^{s\star})$ of man $x$ and woman $y$ when they are single. Given $\theta$ and their chosen allocation $(q^a_{xy},q^b_{xy},Q_{xy})$ when married, we can simply compute the systematic utilities as
\begin{align}\label{eq:systematicutilities}
U_{xy} &= \mathcal{U}^{\theta}_{xy}(q^a_{xy},Q_{xy})-\mathcal{U}_{x0}^{\theta}(q_x^{s\star}, Q_x^{s\star}) \text{ and } \\
V_{xy} &= \mathcal{V}^{\theta}_{xy}(q^b_{xy},Q_{xy})-\mathcal{V}^{\theta}_{0y}(q_y^{s\star}, Q_y^{s\star})
\end{align}

\subsection{Estimation method}
To estimate a ITU-logit model, I rely on the simple characterization of the equilibrium in that particular case. Let me introduce $u_x = -\log\mu_{x0}$ and $v_y = -\log\mu_{0y}$. As a reminder, equilibrium is fully characterized by the system of equations
\begin{equation}\label{eq:estim}
\begin{split}
  \exp(-u_x) + \sum_y \exp(-D^{\theta}_{xy}(u_x,v_y)) &= n_x \\
  \exp(-v_y) + \sum_x \exp(-D^{\theta}_{xy}(u_x,v_y)) &= m_y
\end{split}
\end{equation}

Typically, we will look for the value of $\theta$ that maximizes some objective function under the constraints given by equations (\ref{eq:estim}). The objective function, denoted $F$, can be a log-likelihood function as in GKW or moment-based as in \textcite{GayleShephard2019}.

At this point, there are two ways to proceed to estimate $\theta$. A first solution, which is employed in GKW, is to maximize $F$ with respect to $\theta$, but in that case, one must solve for the $(u_x^\star, v_y^\star)$ solution to (\ref{eq:estim}) for each value of the parameters $\theta$. Given $\theta$, $u_x^\star$, and $v_y^\star$, one can then compute the value of the objective function. These steps are repeated until $F$ is maximized. Although there are efficient algorithms to solve for the equilibrium $(u_x,v_y)$ in system (\ref{eq:estim}), the computational cost is extremely high when there are no analytical formula for the distance function\footnote{The IPFP (Iterative Projective Fitting Procedure) algorithm detailed in GKW and \textcite{GalichonKominersWeber2015} can be used to solve system (\ref{eq:estim}). The issue, however, is that whenever the distance function is not known in closed-form, we must evaluate (numerically) the distance function many times while solving system (\ref{eq:estim}). This is computationally very intensive.}.

A second approach, that I suggest to use here, is to solve the following program:
 \[
 \max_{\theta,u_x,v_y} F(\theta,u_x,v_y)
 \]
 subject to
 \begin{equation*}
\begin{split}
  \exp(-u_x) + \sum_y \exp(-D^{\theta}_{xy}(u_x,v_y)) &= n_x \\
  \exp(-v_y) + \sum_x \exp(-D^{\theta}_{xy}(u_x,v_y)) &= m_y
\end{split}
\end{equation*}
 In other words, we maximize the objective function $F$ with respect to $\theta$ \emph{and} $(u_x,v_y)$, under equilibrium constraints (\ref{eq:estim}). This is very similar to the so-called MPEC approach introduced in, e.g., \textcite{SuJudd2012}. In the next section, I will rely on this second approach.

 As for constructing the objective function, note that when solving problem \ref{eq:P1} for a particular $(x,y)$ pair, we will obtain the predicted mass of marriage for this pair of types (that would be $\exp(-z^\star)$), as well as predicted consumption of private and public goods. Assuming that the marriage patterns and household decisions are observed in the data, then it is easy to construct the objective function.

\section{Application}\label{sec:proofofconcept}
\newcommand{\Model}{Base}
\newcommand{\ModelData}{psid}
\newcommand{\ModelDataSample}{A}

\newcommand{\thepath}{C:/Simon/Dropbox/research/Simon/MatchingCollectiveModels}

In this section, I provide an application of the methods introduced in this paper. In particular, I propose to study the evolution of the sharing rule in the United States since the 1969 as well as housework sharing within households. To do so, I consider a collective model in which the public good is produced by the partners by combining housework time. I use the PSID to construct representative marriage markets in the United States since 1969 (I construct one marriage market roughly every 10 years). I build on an earlier note of \textcite{Weber2016} in which I used my model in a pure simulation exercise. Here, I choose a very parsimonious model that I bring to real data. Therefore, preference parameters estimates should be taken very cautiously. The purpose of the application is twofold. First, it acts as a proof-of-concept and shows how the sharing rule and its distribution can be recovered, when in general in the collective model literature it is only possible to identify the sharing rule up to a constant using distribution factors. Second, the model is structural and allows for counterfactual experiments useful for policy recommendations, which in general is not possible with a reduced-form approach. I study the impact of changes in the gender wage gap on sharing rules and the sharing of housework.

\subsection{Parametric specification}
I shall now be more specific regarding the parameterization of the systematic utilities that appear in equation (\ref{eq:systematicutilities}). First, I assume that singles and couples derive utility from the consumption of a private composite good, $c$, and leisure, $\ell$ (so that $q=(c,\ell)$ in previous notations), and the public consumption of a home-produced good, $Q$. I normalize the price of the private composite good to $1$. All agents can spend time on the labor market, in which case they earn a hourly wage denoted $w$. The total time endowment is $T$.

For singles, the public good is produced from time spent on housework, $h$, according to the production function
 \[
 Q = \tilde{\zeta} h
 \]
Consequently, a single man of type $x\in \mathcal{X}$ faces the following maximization program
\begin{align*}
\max_{c_x^s, \ell_x^s, h_x^s} a_{e(x)}\log c_x^s + \alpha_{e(x)} \log \ell_x^s + A_{e(x)}\log \tilde{\zeta}h_x^s \\
\text{s.t } c_x^s + (\ell_x^s + h_x^s) w_x &\leq Tw_x \\
\ell_x^s + h_x^s &\leq T
\end{align*}
The indices $e(x)$ indicate that I allow preference parameters to vary with types. More precisely, they may depend on a subset of the observable characteristics used to form types. In this particular model, preference parameters vary with the level of education (see below for more details). In similar fashion, a single woman of type $y\in \mathcal{Y}$ solves
\begin{align*}
\max_{c_y^s, \ell_y^s, h_y^s} b_{e(y)}\log c_y^s + \beta_{e(y)} \log \ell_y^s + A_{e(x)}\log \tilde{\zeta}h_y^s\\
\text{s.t } c_y^s + (\ell_y^s + h_y^s) w_y &\leq Tw_y \\
\ell_y^s + h_y^s &\leq T
\end{align*}
It is straightforward to find the optimal $(c_x^{s\star}, \ell_x^{s\star}, h_x^{s\star})$ and $(c_y^{s\star}, \ell_y^{s\star}, h_y^{s\star})$ from which we can compute the singlehood reservation utilities $\mathcal{U}_{x0}^{\theta}(c_x^{s\star}, \ell_x^{s\star}, h_x^{s\star})$ and $\mathcal{V}^{\theta}_{0y}(c_y^{s\star}, \ell_y^{s\star}, h_y^{s\star})$.

If two individuals choose to marry instead, I assume that they can both invest time in the production of the public good. Importantly, I assume that preferences do not change with marriage, but partners have access to a new technology. This is very similar to the approach taken by \textcite{BrowningChiapporiLewbel2013}. Therefore, married partners receive
\begin{eqnarray}
\mathcal{U}_{xy}(c^a_{xy},\ell^a_{xy},Q_{xy}) = \delta^a_{xy} + a_{e(x)}\log c^a_{xy} + \alpha_{e(x)} \log \ell^a_{xy} + A_{e(x)} \log Q_{xy} \\
\mathcal{V}_{xy}(c^b_{xy},\ell^b_{xy},Q_{xy}) = \delta^b_{xy} + b_{e(y)}\log c^b_{xy} + \beta_{e(y)} \log \ell^b_{xy} + B_{e(y)} \log Q_{xy}
\end{eqnarray}
where
\[
Q_{xy} = \zeta (h^a_{xy})^{\eta} (h^b_{xy})^{(1-\eta)}
\]
and $\eta$ and $1-\eta$ are factor shares. Time spent on housework by the man and the woman are denoted $h^a_{xy}$ and $h^b_{xy}$, respectively. I assume that $\eta$ is the same for all types of households\footnote{Of course, it is possible to allow $\eta$ to depend on the type of the man and the woman.}. Finally, the terms $\delta^1_{xy}$ and $\delta^2_{xy}$ capture non economic gains to marriage. They depend on the observable characteristics of both partners. The budget and time constraints are
\begin{align}
c^a_{xy} + c^b_{xy} + (\ell^a_{xy} + h^a_{xy})w_x + (\ell^b_{xy} + h^b_{xy})w_y  &\leq T(w_x + w_y) \\
\ell^a_{xy} + h^a_{xy} &\leq T \\
\ell^b_{xy} + h^b_{xy} &\leq T
\end{align}

The parameters to estimate are $(a_{e(x)}, \alpha_{e(x)}, A_{e(x)}, b_{e(y)}, \beta_{e(y)}, B_{e(y)}, \eta,\delta^a_{xy}$,$\delta^b_{xy})$\footnote{Note that as soon as the $\delta^a_{xy}$ and $\delta^b_{xy}$ contain a constant, the parameters $\zeta$ and $\tilde{\zeta}$ cannot be separately identified from the constant.}, denoted $\theta$. Given a value of $\theta$ and a choice of $(c^a_{xy}, \ell^a_{xy}, h^a_{xy}, c^b_{xy},\ell^b_{xy}, h^b_{xy} )$ for when man $x$ and woman $y$ are married, we can construct the systematic utilities simply by taking the difference $\mathcal{U}_{xy}(c^a_{xy},\ell^a_{xy},Q_{xy})-\mathcal{U}_{x0}^{\theta}(c_x^{s\star}, \ell_x^{s\star}, h_x^{s\star})$ and $\mathcal{V}_{xy}(c^b_{xy},\ell^b_{xy},Q_{xy})-\mathcal{V}^{\theta}_{0y}(c_y^{s\star}, \ell_y^{s\star}, h_y^{s\star})$.

\subsection{Data}\label{sec:data}
\newcommand{\edone}{BHS}
\newcommand{\edtwo}{HS}
\newcommand{\edthr}{C}
The data used for estimating the model is the PSID (Panel Study of Income Dynamics). The basic panel consists of $5,000$ families (over $18,000$ individuals), and it is representative of the US population. The study was repeated yearly since 1968 and every two years after 1997. In this application, I use only the information available in the core dataset, since we only need basic information on the education level of individuals, their labor supply, time spent on housework, income and marital status.

Using PSID data, I construct six marriage markets (roughly one every ten years), one from each of the following waves: 1969, 1977, 1987, 1997, 2007 and 2017. To construct the samples used in this application, I apply the following selection rules. First, I keep all single men (heads of their household) aged between $42$ and $52$ years old and all single women (heads of their household) aged $40-50$\footnote{These age intervals were chosen because the average age difference between married men and women is two years. This way of proceeding is standard in the literature, see \textcite{ChiapporiSalanieWeiss2017}.}. Second, I select heterosexual couples (legally married or cohabitating and heads of their household) in which both partners satisfy these age requirements. This is similar to \textcite{ChiapporiSalanieWeiss2017} who also use data on older couples to limit the problem of truncation (individuals who are single past 40 are not likely to ever marry). Therefore, my focus is on a more stable marriage market. In addition, there is less variance in the labor supply and housework hours of men and women at these ages, as children (if any) are older. I use information on annual labor supply and labor income to construct hourly wages (abnormally high or low wages are trimmed). When wages are not observed, I predict them using a Heckman selection model \parencite{Heckman1979}\footnote{The second stage estimation includes detailed education dummies, age, age squared and regional dummies. The first stage regression includes the same variables as well as marital status and the presence of children under age of 5 in the household.}. I observe the number of years of education for each individual which allows me to categorize men and women into two broad education types: ``High school and below'' (HS) and ``College'' (C). In addition, I observe how much time men and women spend on housework. Naturally, this variable is a very imperfect measure of time spent producing household goods, since it is self reported and it is hard to know what respondents take into account when providing an answer. We know however, that this includes time spend cooking, cleaning and doing work for the house (core housework). Households with missing information are deleted. Finally, I fix the time endowment to $112$ hours per week and assume that leisure is equal to the total time endowment minus the time spent on the labor market and on housework.

\begin{figure}
  \caption{Trends}\label{fig:trends}
  \centering
\begin{subfigure}[b]{0.35\textwidth}
\caption{Couples}
  \includegraphics[width=0.9\textwidth, trim=0cm 0cm 0cm 0cm,clip]{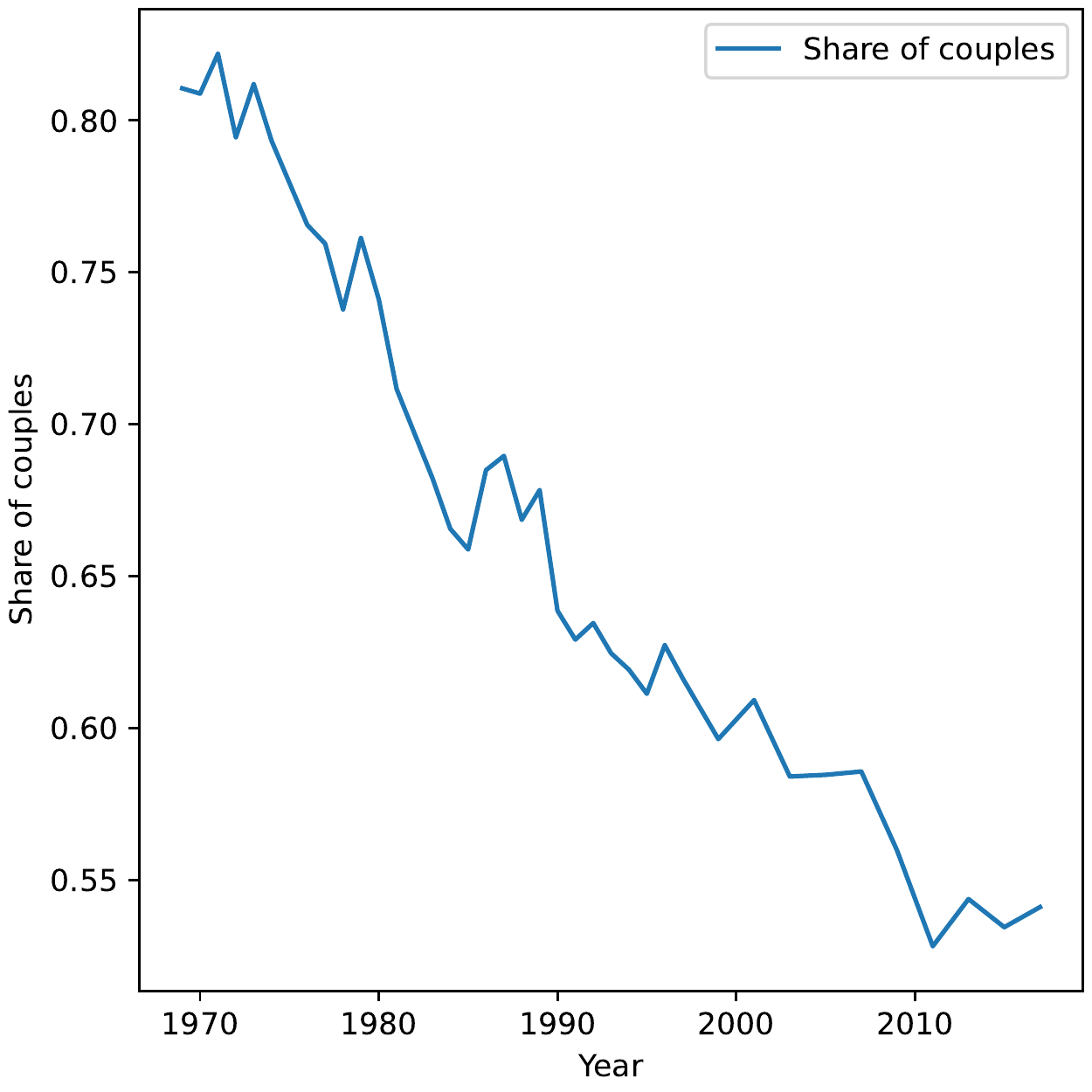}
  \end{subfigure}
  \hspace{1cm}
  \begin{subfigure}[b]{0.35\textwidth}
  \caption{Singles}
  \includegraphics[width=0.9\textwidth, trim=0cm 0cm 0cm 0cm,clip]{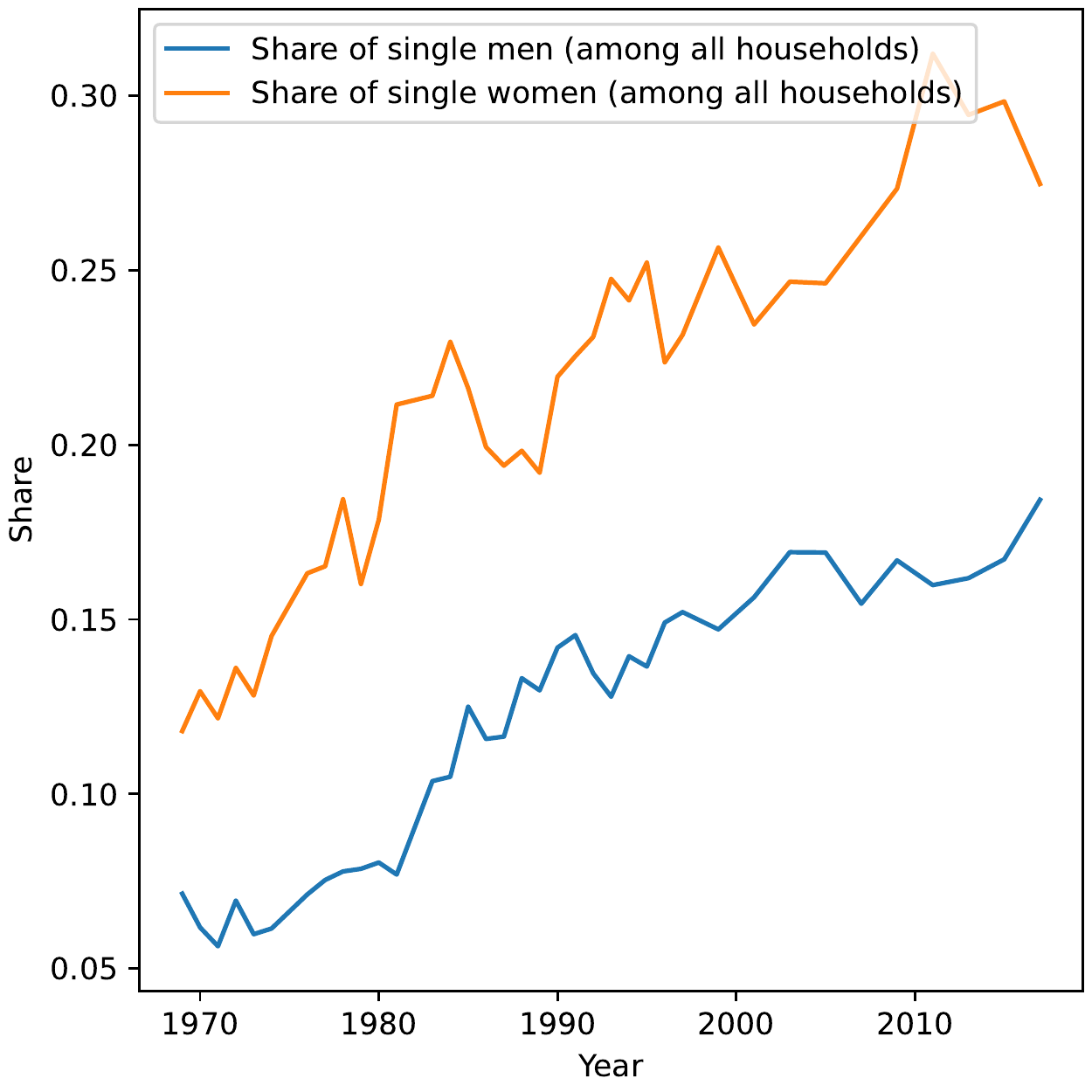}
  \end{subfigure}

  \begin{subfigure}[b]{0.35\textwidth}
\caption{Education}
  \includegraphics[width=0.9\textwidth, trim=0cm 0cm 0cm 0cm,clip]{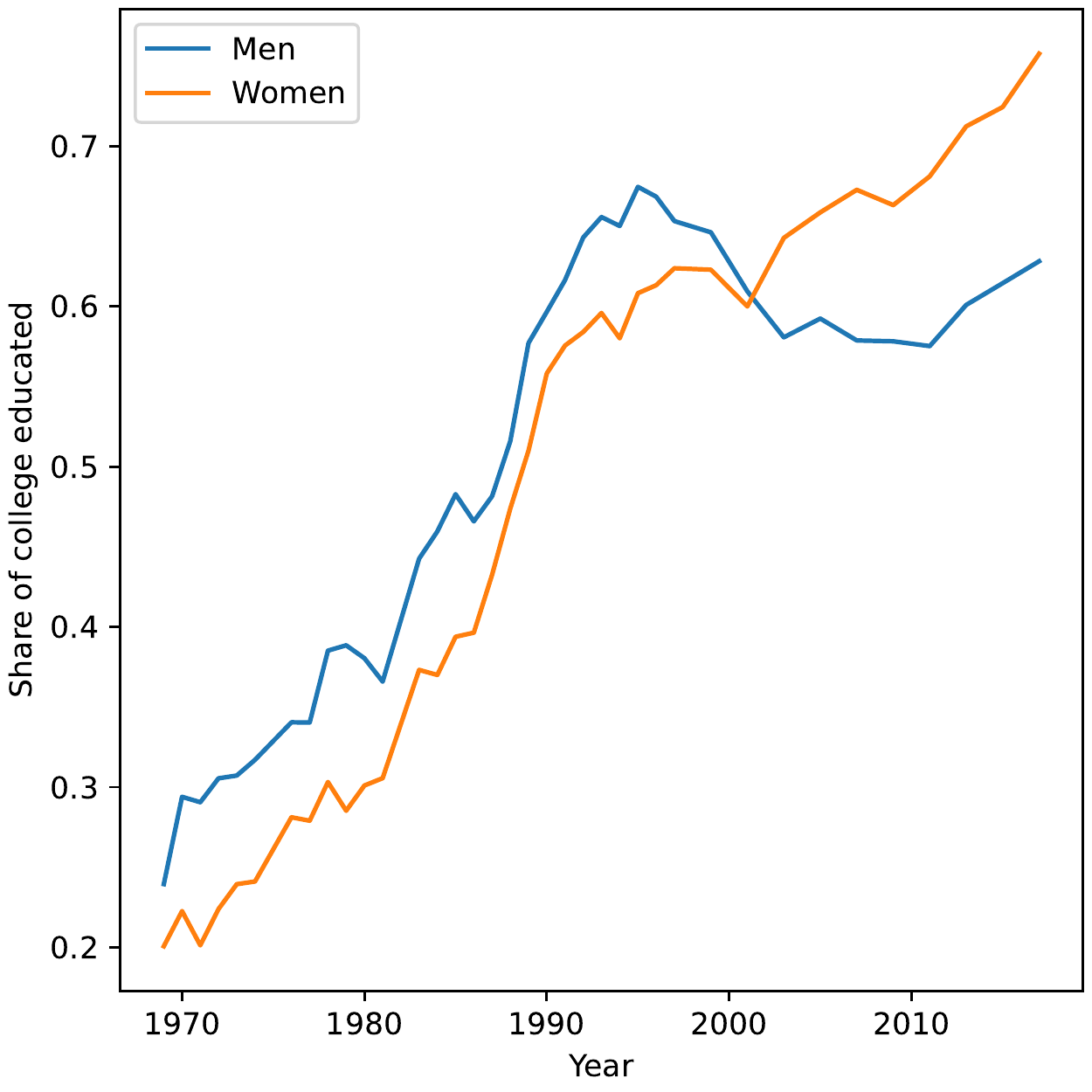}
  \end{subfigure}
  \hspace{1cm}
  \begin{subfigure}[b]{0.35\textwidth}
  \caption{Wages}
  \includegraphics[width=0.9\textwidth, trim=0cm 0cm 0cm 0cm,clip]{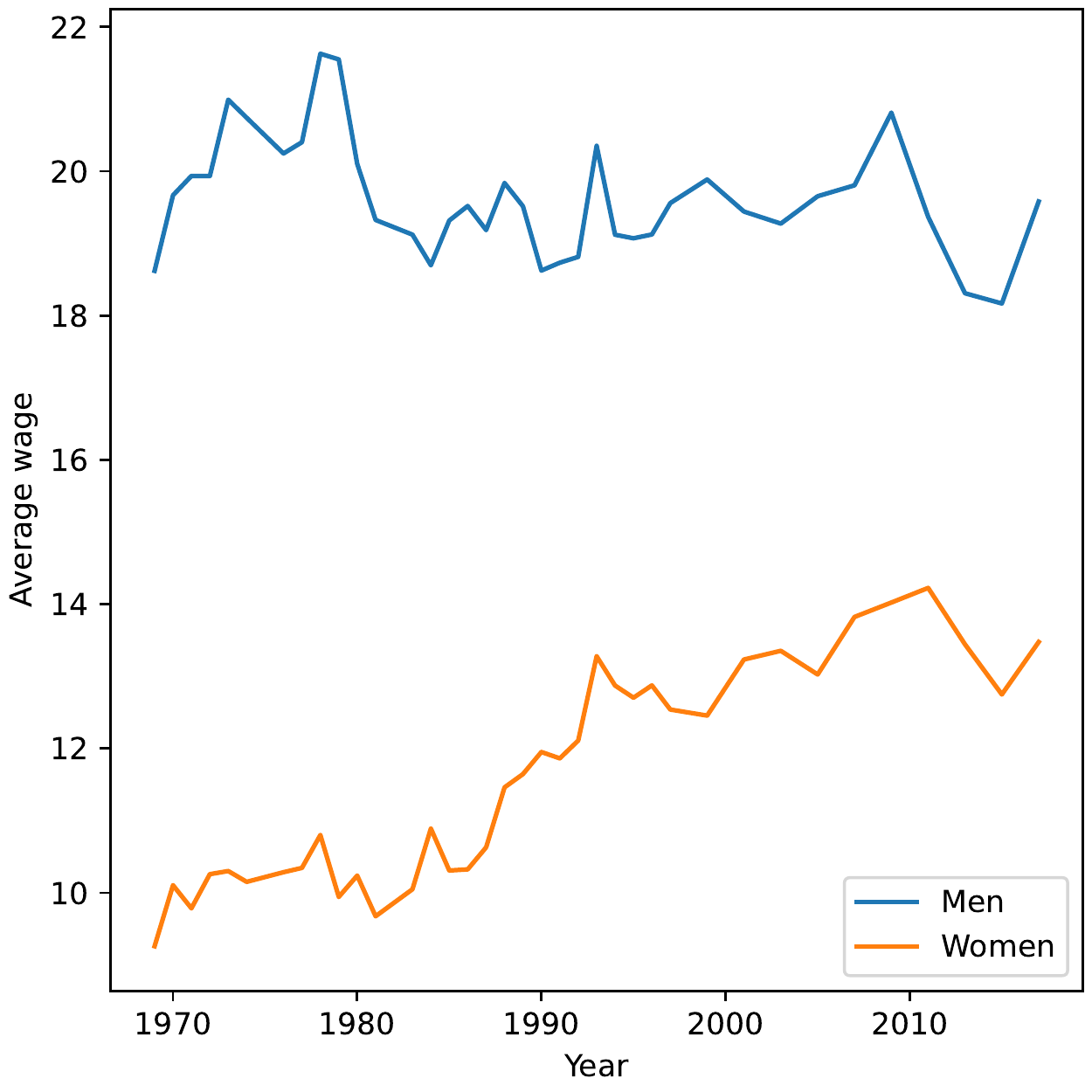}
  \end{subfigure}

  \begin{subfigure}[b]{0.35\textwidth}
\caption{Employment rate}
  \includegraphics[width=0.9\textwidth, trim=0cm 0cm 0cm 0cm,clip]{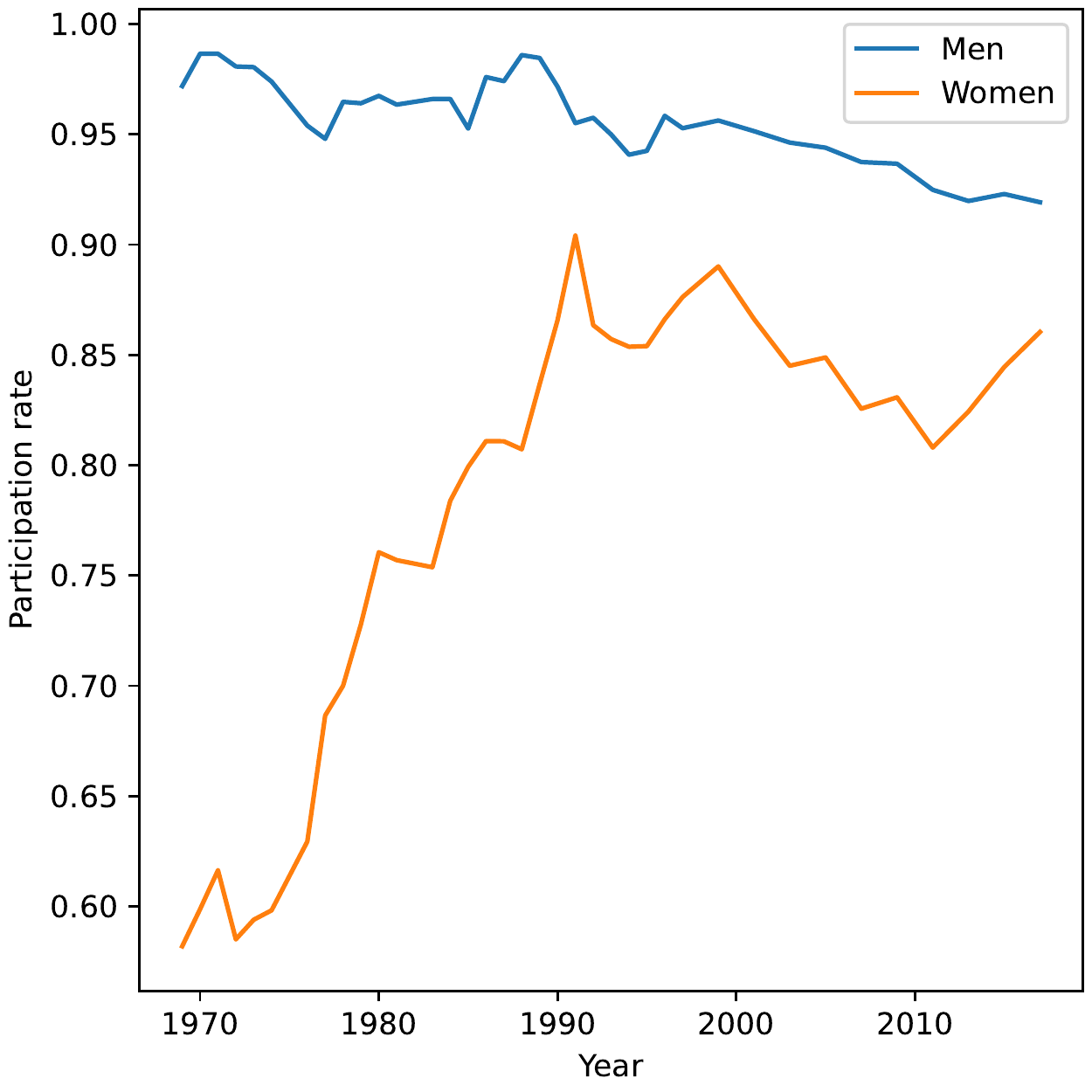}
  \end{subfigure}
  \hspace{1cm}
  \begin{subfigure}[b]{0.35\textwidth}
  \caption{Housework hours}
  \includegraphics[width=0.9\textwidth, trim=0cm 0cm 0cm 0cm,clip]{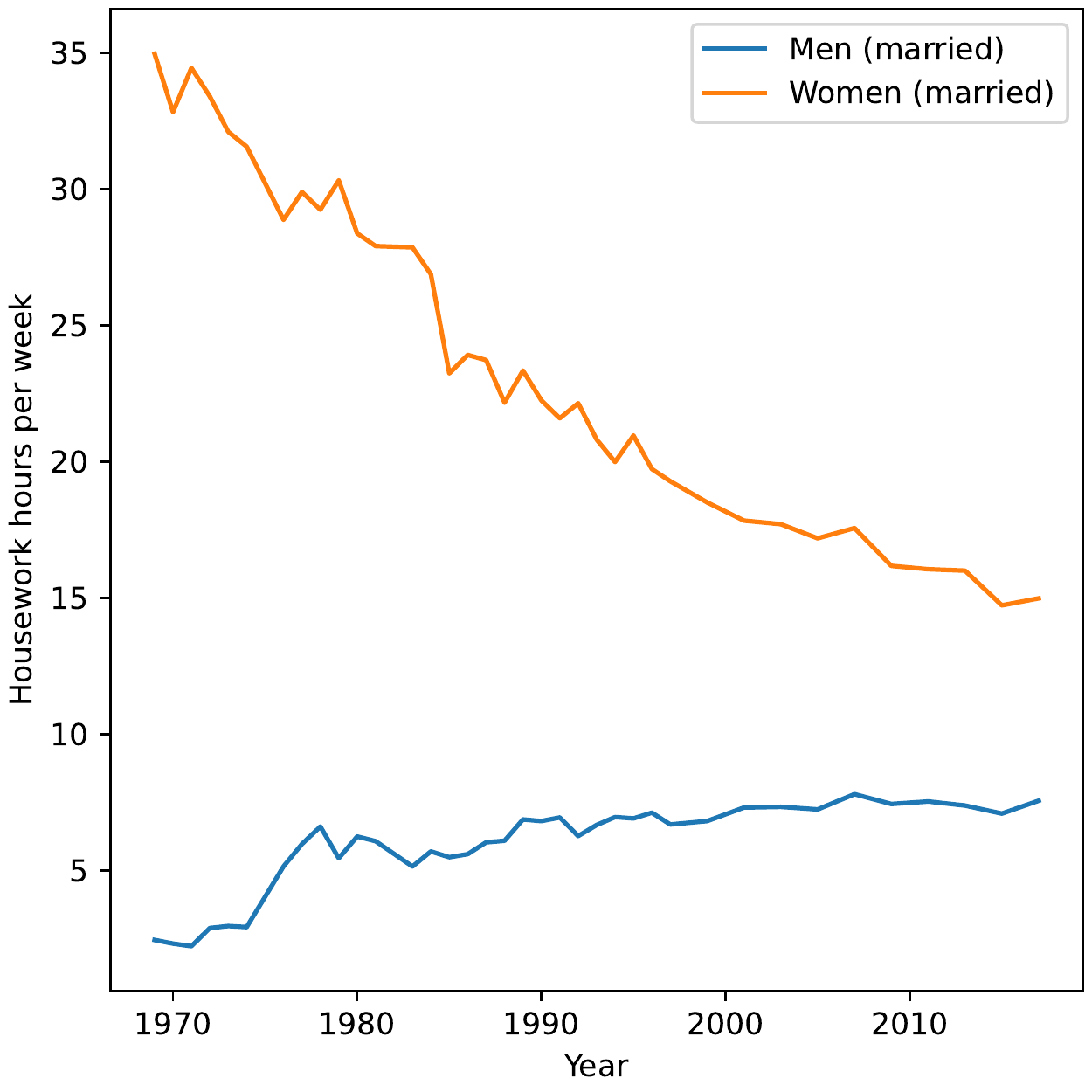}
  \end{subfigure}

  \caption*{\tiny This figure shows trends for selected characteristics and variables from 1969 to 2017: (a) share of couples among all households, (b) share of single men and women among all households, (c) share of college educated individuals (by gender), (d) mean wages (by gender), (e) employment rate (by gender), (f) mean hours of housework per week (by gender) among individuals living in a couple.}
\end{figure}

I report summary statistics on the important variables in table \ref{tab:descstats} for three selected years: 1977, 1997 and 2017\footnote{Summary statistics for the other three marriage markets are available upon request.}. Figure \ref{fig:trends} show trends in the following key variables: the share of couples, single men and single women among the selected households (panels (a) and (b)), the share of college educated men and women (panel (c)), average wages by gender (panel (d)), the employment rate by gender (panel (e)), and the average number of hours spent on housework per week by individuals living in couples (panel (f)).

\begin{table}
\caption{Descriptive statistics}\label{tab:descstats}
\begin{tabular}{L{5cm}cccccc}
\toprule
& \multicolumn{2}{c}{1977} & \multicolumn{2}{c}{1997} & \multicolumn{2}{c}{2017}\\
& Women & Men & Women & Men & Women & Men \\
\midrule
\textbf{A. Couples} &&&&&&\\
Age &44.756 &47.294 &44.849 &46.747 &44.703 &46.477 \\ 
Wage &10.055 &20.793 &12.683 &20.339 &14.071 &21.038 \\ 
Hours of work &16.776 &42.428 &28.075 &42.447 &30.620 &41.350 \\ 
Hours of housework &29.905 &5.981 &19.290 &6.699 &14.994 &7.570 \\ 
Share college-educated &0.286 &0.340 &0.624 &0.655 &0.780 &0.656 \\ 

&&&&&&\\
\textbf{B. Singles} &&&&&&\\
Age &45.228 &47.192 &44.756 &46.133 &44.402 &46.400 \\ 
Wage &11.677 &16.461 &12.158 &16.398 &12.303 &15.301 \\ 
Hours of work &25.741 &32.124 &33.221 &37.272 &33.169 &35.540 \\ 
Hours of housework &22.421 &6.423 &11.276 &7.106 &11.598 &8.573 \\ 
Share college-educated &0.246 &0.346 &0.622 &0.646 &0.713 &0.545 \\ 
\\
\bottomrule
\end{tabular}
\caption*{\raggedright\tiny Note: this table shows summary statistics for three marriage markets (1977, 1997, 2017), constructed using the selection rules mentioned in the main text. In particular, I keep men aged 42-52 and women aged 40-50.}
\end{table}

On average, married men are two years older than their wife (this is true even without applying the age selection criteria) and this has barely changed over time. The share of couples has steadily declined over the period while the share of singles has mechanically increased. The marriage market is unbalanced, as men are more scarce than women. This comes mainly from the fact that women are over-represented among the divorced and separated as they find it more difficult to enter a new marriage (see chapter 1 in \textcite{BrowningChiapporiWeiss2014}). Table \ref{tab:descstats} and figure \ref{fig:trends} show important changes in the educational attainment of men and women. While a majority of them did not go to college in 1969, a majority has at least some college education in 2017. Women also appear now to be higher educated than men (this fact is discussed in \textcite{ChiapporiSalanieWeiss2017}). Wages (expressed in 1999 dollars) have remained fairly stable for men over time, while the wages of women have dramatically increased, reducing the gender wage gap. The share of men working a strictly positive number of hours (what I refer to as employment rate) and their average hours of work per week have been quite stable over the period. For women, these have dramatically increased since 1969, although they have stabilized since the 1990s. Finally, men spend much less time on housework than women, but the sharing of housework time is much more balanced in 2017 than it was in 1969. These gaps narrow when comparing single men and single women instead of married men and married women. Figure \ref{fig:trends} shows the dramatic decline in the number of hours of housework spent by women living in a couple, while in the same period time spent by men has increased. Despite these changes, women still take care of a substantial share of housework time ($95\%$ in 1969 and $65\%$ in 2017). Finally, in appendix \ref{app:results}, I tabulate the observed matching by education level, which suggest strong assortative matching in education.

\subsection{Identification and estimation}
To estimate the model, I follow the steps described in section \ref{sec:estimation}. I assume that there is a mass $1$ of each
type of man and woman in the sample\footnote{Here, I could adapt the model to the continuous logit framework a-la-\textcite{DupuyGalichon2014}, but that does not change anything to the estimation, as pointed out in \textcite{GalichonKominersWeber2019}.}, that is $n_x = m_y = 1$ for all $x, y \in \mathcal{X} \times \mathcal{Y}$. Preference parameters are allowed to vary with education, as explained above.

The parameters of interest, $(a_{e(x)}, \alpha_{e(x)}, A_{e(x)}, b_{e(y)}, \beta_{e(y)}, B_{e(y)}, \eta)$, are identified from the sole observation of the marriage patterns, labor supplies, time spent on housework and wages. First, I impose for simplicity that the parameters $a$, $\alpha$ and $A$ sum to $1$. Next, in a collective model with a home-produced nonmarketable good, inputs choice is driven by cost minimization. Therefore, the ratio $\eta/(1-\eta)$ is equal to the ratio of wages ($w_x/w_y$) times the ratio of housework ($h^a_{xy}/h^b_{xy}$), which shows identification of $\eta$. The identification of the other parameters proceeds as follows. First, note that observing private consumption is not necessary\footnote{In most available survey data, private consumption is poorly measured. At best, only the aggregate private consumption of the couple is known. Whenever private consumption is well measured, there is no need for collective models, see e.g. \textcite{BrowningBonke2006}.} for identification since efficiency (abstracting from corner solutions) imposes $c^a_{xy} = (a_{e(x)}/\alpha_{e(x)})w_x \ell^a_{xy}$. Second, we can combine these efficiency restrictions, the systematic utilities identified from the marriage patterns and the program for singles to identify the other parameters.\footnote{As discussed above, the systematic utilities of men and women also contain a non-economic component, $\delta^a_{xy}$ and $\delta^b_{xy}$. I parametrize this component as a linear function of partner's characteristics. The included characteristics are: dummies for each education level, dummies for each possible level of the absolute value of the difference between the education level of the partners, age of the partners, and absolute value of the age difference. The parameters associated to these characteristics are obviously parametrically identified.}

The objective function used to estimate $\theta$ is the log-likelihood constructed from the predicted frequency that each pair will form on the marriage market, as well as the predicted labor supply and housework time. Appendix \ref{app:estim} contains details about the log-likelihood. Finally, I estimate the model on a subsample of 400 households randomly drawn from the full sample described in table \ref{tab:descstats}\footnote{Using more observations is feasible. The model can be estimated in a reasonable amount of time with up to 1000 households.}. The subsample contains approximatively 200 couples, which is roughly equivalent to what can be found in other studies such as \textcite{CherchyeDeRockVermeulen2012a} and \textcite{DelBocaFlinn2014}.

\subsection{Results}\label{sec:results}
\subsubsection{Preference parameters and model fit}
To estimate the model, I use the observed marriage patterns and labor supplies and housework time of married couples and singles. To provide a better fit of the marriage patterns, the model's $\delta^a_{xy}$ and $\delta^b_{xy}$ terms includes dummies for each education type as well as a dummy when the man and the woman have the same education level. They also depend linearly on the age of the partners and the absolute value of the difference in age. Economically speaking, these components capture costs (or gains) of singlehood, as well as preferences for education and age assortativeness\footnote{These two traits are indeed the most important ones for explaining marriage patterns. See, e.g., \textcite{CiscatoWeber2020} for the US marriage market, or \textcite{DoorleyDupuyWeber2019} for Germany.}.

Estimates for the main parameters for selected years are displayed in table \ref{tab:estimates}\footnote{I only present the estimates for the parameters $(a_{e(x)}, \alpha_{e(x)}, A_{e(x)}, b_{e(y)}, \beta_{e(y)}, B_{e(y)}, \eta)$ since these are the most important ones. Not surprisingly, the other non-economic gains parameters suggest a taste for assortative mating in education and age. These estimates are available upon request.}. It is difficult to find clear patterns in the estimates, however women do seem to care more about public consumption, with only slight differences across types. The estimate for $\eta$ is about $0.10$ in 1969 and $0.4$ in 2017. The magnitude of this parameter is not surprising: e.g. for 2017, the ratio $\eta/(1-\eta)$ should be equal to the ratio of wages ($w_x/w_y$) times the ratio of housework ($h^a_{xy}/h^b_{xy}$). From the raw data, we can deduce that $\eta$ should be around $0.40$. The same computations can be performed for the other years.

The model fit is discussed in appendix \ref{app:results} in table \ref{tab:modelfit}. The model fits the data reasonably well, considering that the model is very parsimonious. The table shows the mean of predicted and observed hours of work and housework of men and women for selected marriage markets.

\FloatBarrier
\renewcommand{\arraystretch}{1}
\begin{table}[h]
\centering
\caption{Parameter Estimates}\label{tab:estimates}
\begin{threeparttable}
\begin{tabular}{cccc|ccc|c}
\toprule
  & \multicolumn{3}{c}{Men} & \multicolumn{3}{c}{Women} &\\
  \cline{2-4}
  \cline{5-8}
  & Private C. & Leisure & Public C. & Private C. & Leisure & Public C. & Technology \\
    & $a$ & $\alpha$ & $A$ & $b$ & $\beta$ & $B$ & $\eta$ \\

\textbf{A. 1977.} & &  &  & &  &  & \\
HS &0.309 &0.648 &0.043 &0.255 &0.526 &0.220 &0.218 \\ 
C &0.316 &0.646 &0.039 &0.316 &0.498 &0.187 &0.218 \\

& &  &  & &  &  & \\
\textbf{B. 1997.} & &  &  & &  &  & \\
HS &0.323 &0.640 &0.036 &0.302 &0.562 &0.136 &0.326 \\ 
C &0.369 &0.584 &0.047 &0.314 &0.560 &0.127 &0.326 \\

& &  &  & &  &  & \\
\textbf{C. 2017.} & &  &  & &  &  & \\
HS &0.314 &0.616 &0.070 &0.251 &0.634 &0.116 &0.433 \\ 
C &0.372 &0.571 &0.057 &0.336 &0.566 &0.098 &0.433
\\
\bottomrule
\end{tabular}
\vspace{-0.5ex}
\begin{tablenotes}
\item \linespread{1.0}\footnotesize Note: types are displayed in rows (HS = high school and below, and C = college). Estimates are presented for three selected marriage markets: 1977, 1997 and 2017.
\end{tablenotes}
\end{threeparttable}
\end{table}
\FloatBarrier

\subsubsection{Sharing rules}
The model predicts Pareto weights for each household. However, I choose to report a more informative, alternative, monetary measure of power: the sharing rule. In the collective model literature, sharing rules are in general preferred since their interpretation is easier: they simply express how resources (a certain amount of money) are divided among partners. For a particular pair $(x,y)$, denote $\hat{c}^a_{xy}$, $\hat{c}^b_{xy}$, $\hat{\ell}^a_{xy}$ and $\hat{\ell}^b_{xy}$ the predicted private consumption and leisure of the man and the woman. Similarly, introduce $\hat{Q}_{xy}$ the predicted consumption of public good. The conditional sharing rule (from the point of view of women) is computed as follow
\[
S^{cond}_{xy} = \frac{\hat{c}^b_{xy} + w_y \hat{\ell}^b_{xy}}{\hat{c}^a_{xy} + w_y \hat{\ell}^a_{xy} + \hat{c}^b_{xy} + w_y \hat{\ell}^b_{xy}}
\]
while the (unconditional) sharing rule is
\[
S_{xy} = \frac{\hat{c}^b_{xy} + w_y \hat{\ell}^b_{xy} + P^{b,a}_{xy}\hat{h}^a_{xy} + P^{b,b}_{xy}\hat{h}^b_{xy}}{\hat{c}^a_{xy} + w_x (\hat{\ell}^a_{xy} + \hat{h}^a_{xy})  + \hat{c}^b_{xy} + w_y (\hat{\ell}^b_{xy} + \hat{h}^b_{xy})}
\]

\begin{figure}
  \caption{Sharing Rule}\label{fig:sharingrule}
  \centering
\includegraphics[width=0.7\textwidth, trim=0cm 0cm 0cm 0cm, clip]{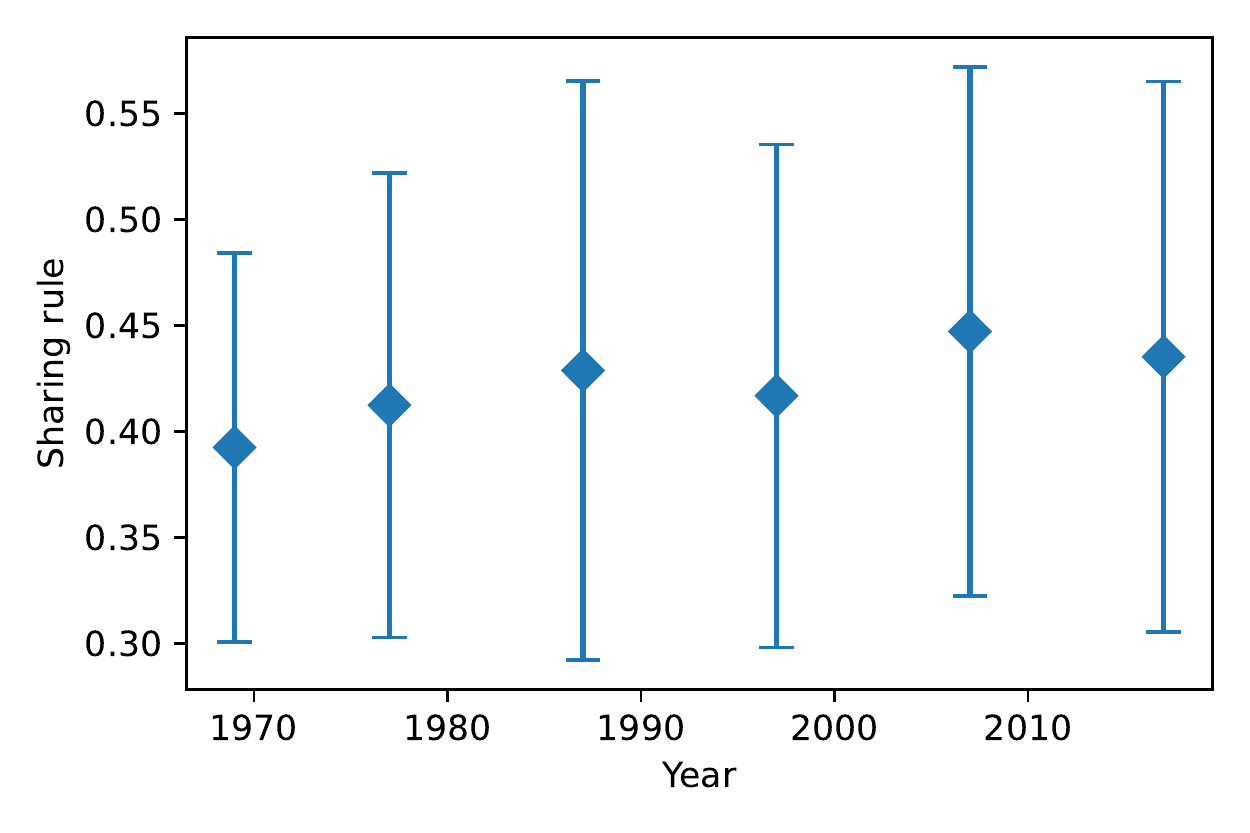}
  \caption*{\raggedright\tiny This figure shows the distribution of the sharing rule (the share of women's consumption in total consumption) over all observed couples for each of the six marriage markets: 1969, 1977, 1987, 1997, 2007 and 2017. The diamond marker indicates the mean unconditional sharing rule while the bottom and top caps indicate the first and third quartile of the distribution of the unconditional sharing rule, respectively.}
\end{figure}

Because the model is parametric, it is possible to recover the Lindhal prices $P^{b,a}_{xy}$ and $P^{b,b}_{xy}$ for the woman (where $w_y = P^{a,b}_{xy}+P^{b,b}_{xy}$) associated with the public goods $h^a$ and $h^b$\footnote{In this application with log-Cobb Douglas utility functions and Cobb Douglas production function, the hours of housework $h_{xy}^a$ and $h_{xy}^b$ can be seen as two distinct public goods that can be purchased at price $w_x$ and $w_y$, respectively. The Lindhal prices are given by the first order conditions of the decentralized utility maximization problem. For example, for the woman the prices are $P^{b,a}_{xy} = (\partial \mathcal{V}_{xy}/\partial c_{xy}^b)/(\partial \mathcal{V}_{xy}/\partial h_{xy}^a)$ and $P^{b,b}_{xy} = (\partial \mathcal{V}_{xy}/\partial c_{xy}^b)/(\partial \mathcal{V}_{xy}/\partial h_{xy}^b)$.}. Therefore, we can compute the unconditional sharing rule, whose distribution is showed in figure \ref{fig:sharingrule} for all six marriage markets since 1969 (for the couples that are observed in the data). Overall, it seems that women are less favoured than men, a prediction that can be explained by the more attractive outside options of men (since they have a higher wage), and the population imbalance (the sex ratio favours men). However, women fare better in 2017 than in 1969: the mean unconditional sharing rule has gradually increased from about 0.39 in 1969 to 0.435 in 2017. Note that the figure shows the unconditional sharing rule, which usually sits slightly above the conditional sharing rule since all households produce and consume at least some public good, and that women tend to value it more.

These results are in line with the literature. In terms of magnitude, several papers seem to suggest that the (mean) sharing rule is quite close to 0.5, but is slightly less favourable to women. For example, \textcite{LiseSeitz2011} obtain a mean sharing rule of 0.386. That paper produces estimates of the sharing rule by bringing a mixed logit model to consumption data in the UK from 1968 to 2001. \textcite{CherchyeDemuynckDeRockEtAl2017}, who use a revealed preferences approach and stability conditions on the marriage market to obtain tight bounds for the sharing rule (using recent Dutch data), seem to support this conclusion as well. Few papers have looked at the evolution of the sharing rule over time. Among them, \textcite{LiseSeitz2011} suggest that the sharing rule has increased over time.

\begin{figure}
  \caption{Sharing Rule, by education of partners}\label{fig:sharingrulebyeduc2017}
  \centering
\includegraphics[width=0.7\textwidth, trim=0cm 0cm 0cm 0cm, clip]{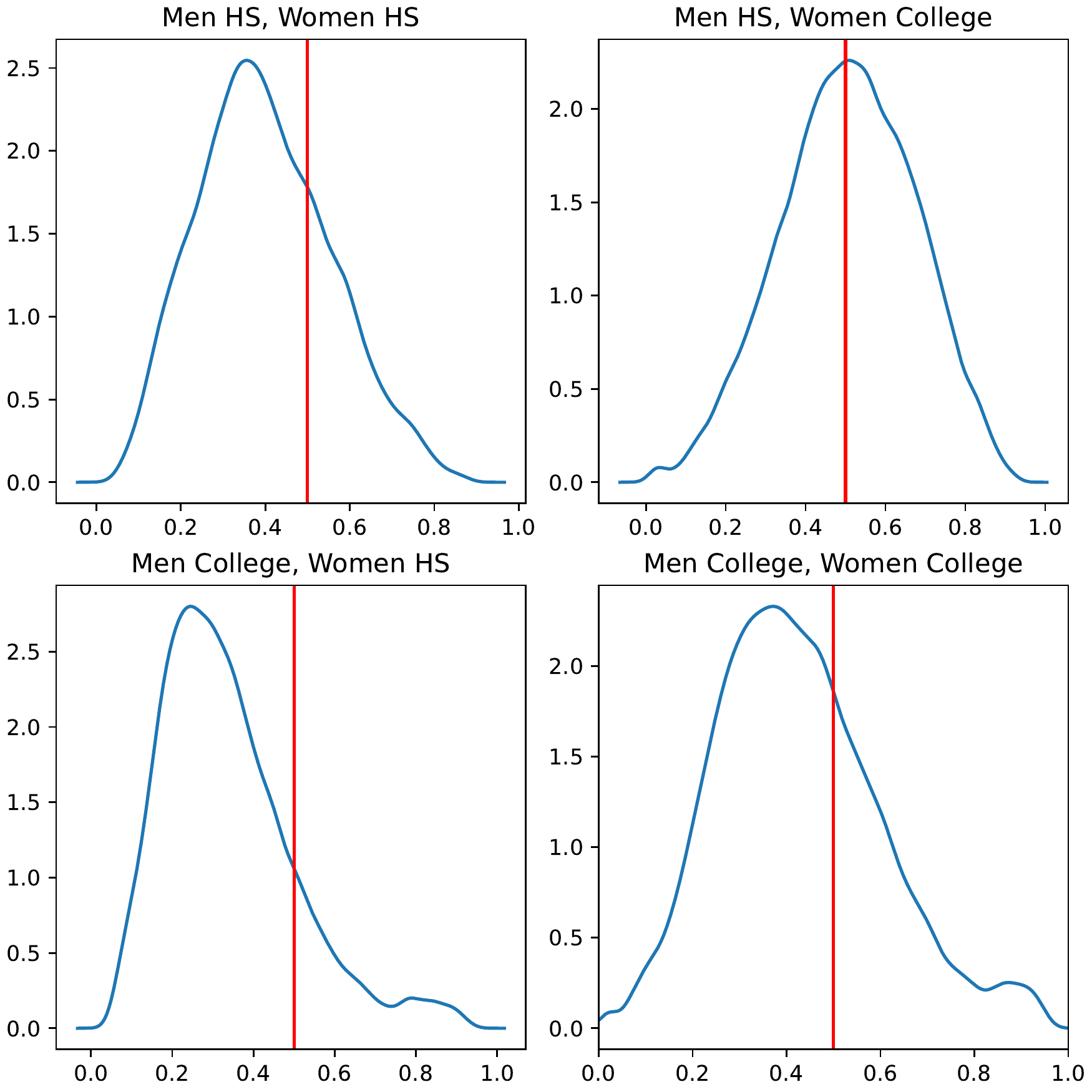}
  \caption*{\raggedright\tiny This figure shows the distribution of the sharing rule in 2017 over all $(i,j)$ pairs, conditioning on the education levels of the husband and the wife. The husband education level is displayed in rows, the one of the wife in columns.}
\end{figure}

Sharing rules can be estimated for all possible couples in each marriage market. In figure \ref{fig:sharingrulebyeduc2017}, I choose to focus on 2017 and plot the density of the sharing rules conditional on the education level of the partners. Men's education is displayed in rows and women's education is represented in columns. The estimates are well behaved in the sense that the distributions are shifting to the right whenever the education of women increases and shifting to the left whenever the education of men increases. This is consistent with the fact that outside options improve with education (since higher educated individuals are likely to have higher wages). This is very much in line with the findings of \textcite{ChiapporiCostaDiasMeghir2018}, although their structural matching model features transferable utility.

\subsubsection{Counterfactual experiment: no gender wage gap}
In this paragraph, I proceed with exploring a counterfactual corresponding to a decrease in the gender wage gap. I focus on the most recent marriage market in my data (2017) and increase the wage of all women in the same proportion, up until the point there is no longer any gender wage gap, on average. Since this improves the outside option of women of staying single, we can expect an improvement in the bargaining power of women. After increasing the wage of women, I compute the new marriage market equilibrium by solving system (\ref{eq:estim}) using the IPFP algorithm (see GKW). Preferences are assumed to be unchanged in this new market.

\begin{figure}[h]
  \caption{Counterfactual sharing rule and housework sharing}\label{fig:counter1}
  \centering
\begin{subfigure}[b]{0.4\textwidth}
  \includegraphics[width=\textwidth, trim=0cm 0cm 0cm 0cm,clip]{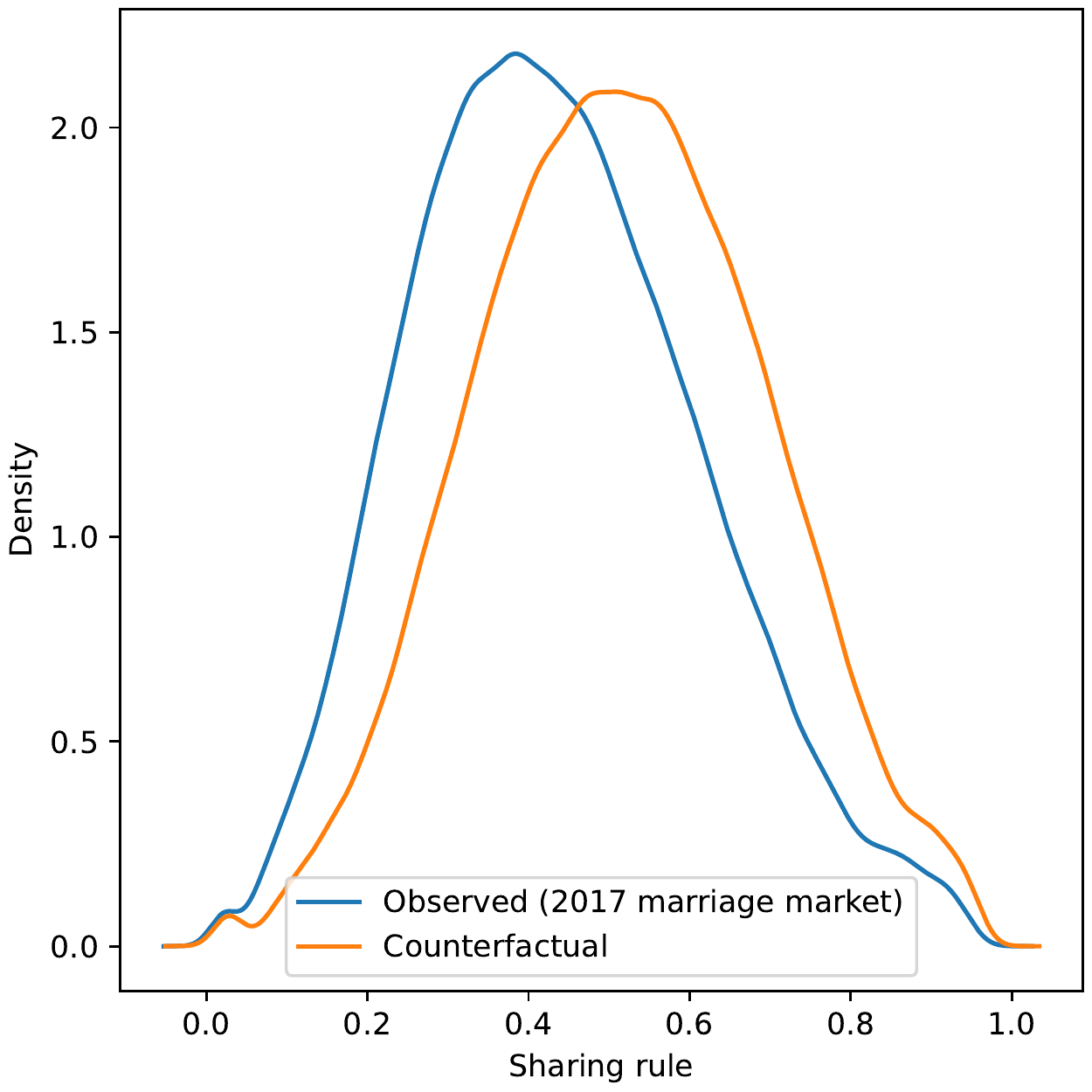}
  \end{subfigure}
  \begin{subfigure}[b]{0.4\textwidth}
  \includegraphics[width=\textwidth, trim=0cm 0cm 0cm 0cm,clip]{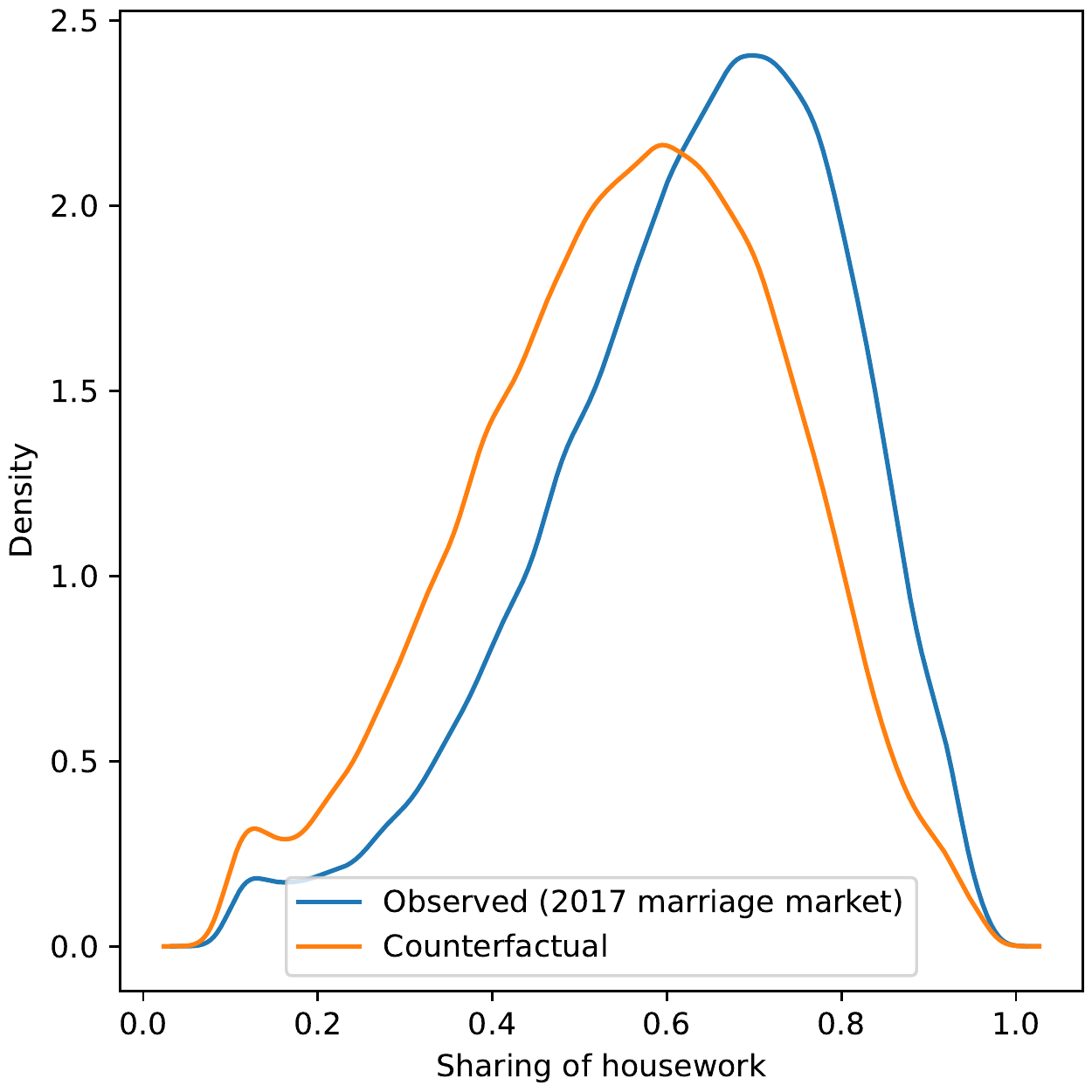}
  \end{subfigure}
  \caption*{\tiny This figure shows the distribution of the sharing rule and the sharing of housework time in the 2017 marriage market and in the counterfactual marriage market. The counterfactual corresponds to a closing of the gender wage gap.}
\end{figure}

The results are presented in figure \ref{fig:counter1}. The left panel shows the distribution of the unconditional sharing rule as predicted by the model in 2017 and in the counterfactual. As expected, as women earn more, the sharing rule shifts to the right in favour of women. The mean sharing rule increases from 0.435 to 0.514. The right panel displays the distribution of the sharing of housework time in the household. As their wages and bargaining power increases, women spend more time on the labor market, less time on housework (and they take care of $55\%$ of total housework time under the counterfactual, as opposed to $63\%$ in the observed case), and enjoy more leisure.

\subsubsection{Discussion.}
In our ITU framework, public good production responds to changes in the woman's wage because of two effects: a direct effect (due to the change in the price of an input, i.e. the woman's time) and an indirect effect (due to the shift in bargaining power). In a TU framework, only the direct effect would apply.

\begin{figure}
  \caption{Public good production}\label{fig:houseworksharing}
  \centering
\includegraphics[width=0.6\textwidth, trim=0cm 0cm 0cm 0cm, clip]{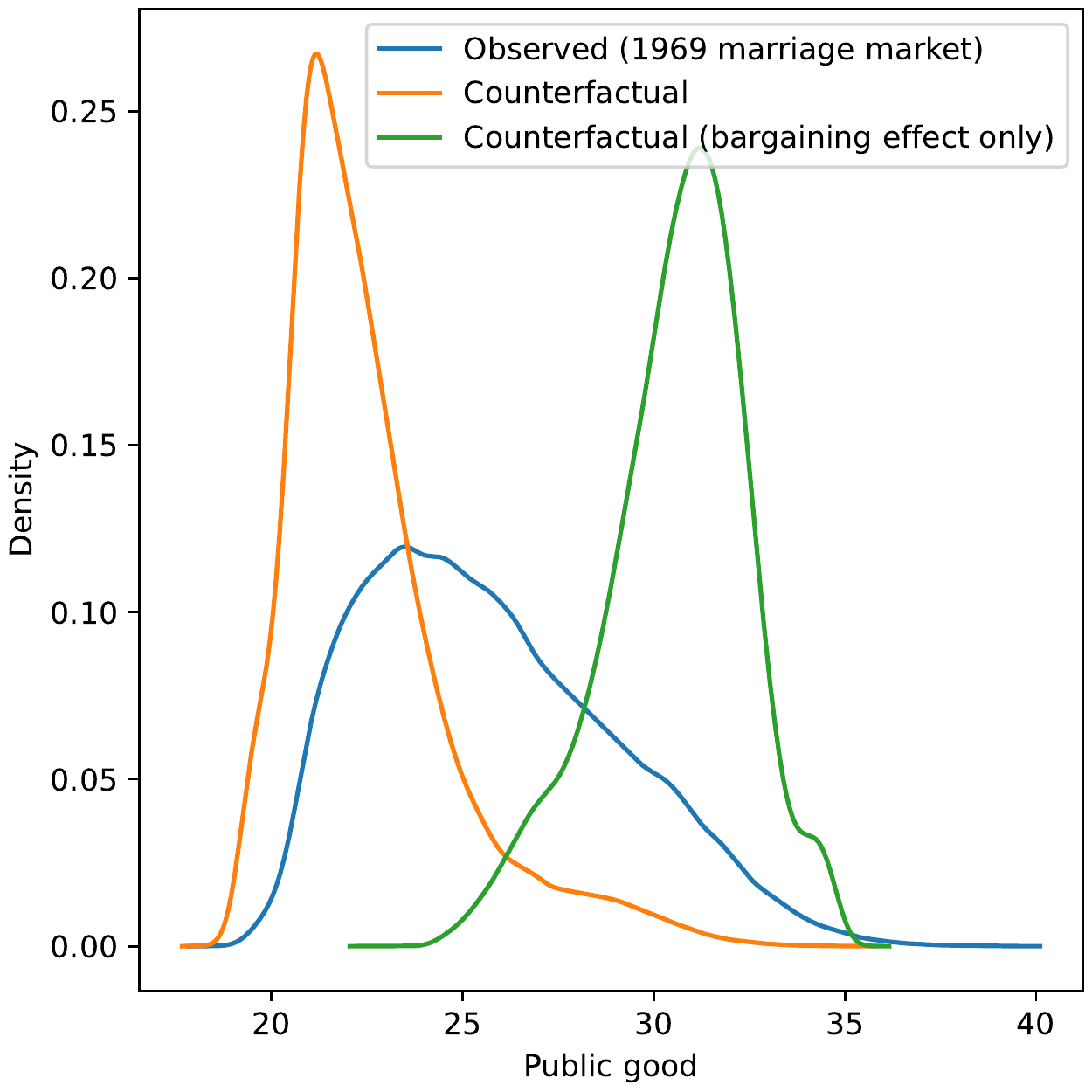}
  \caption*{\raggedright\tiny This figure shows the distribution public good production over all $(i,j)$ pairs in the 1969 marriage market, the counterfactual market (1969 market with no gender wage gap), and the counterfactual where only the effect of the shift in bargaining power is shown.}
\end{figure}

Our methodology is well suited to decompose these effects. Recall that Pareto weights are obtained as the Lagrange multipliers associated to the utility constraints in the distance function optimization problem (see theorem \ref{thm:collectivemodels}). To isolate the effect of a shift in bargaining power on household decisions from changes in prices or income, we can simply solve the collective model using the counterfactual Pareto weights while keeping prices and incomes fixed to their observed values. I illustrate this approach with a second counterfactual experiment: this time I focus on the 1969 marriage market\footnote{I choose to focus on the 1969 marriage market for illustrative purposes, because the men's and women's preferences are more markedly different in earlier markets.} and change the wages of women so that there is no longer any gender wage
gap, on average. Figure \ref{fig:houseworksharing} shows the distribution of public good production as predicted by the model in the 1969 marriage market and in the counterfactual (1969 marriage market with no gender wage gap). It also shows the distribution of public good production had wages not been changed, but had bargaining power shifted as in the counterfactual. As expected, the shift in bargaining power results in an increase in public good production, since women value the good more than men. However, the increase in the price of one of the inputs (women's time) is so massive that in the counterfactual the production of public good decreases.

Note that there are many other counterfactual experiments that could be designed\footnote{I also considered a counterfactual experiment in which the 2017 gender wage gap is increased to its 1969
level. I find that the average sharing rule would fall from 0.435 to 0.365 and the sharing of housework would increase from $63\%$ to $70\%$. These additional results are available upon request.}. For example, one could increase the
number of men on the marriage market (by increasing their mass) so that the sex ratio becomes more and more balanced. As men become less scarce, we expect an increase in women's bargaining power. One advantage of the present framework is that such distribution factors (wage ratio, sex ratio, etc) arise naturally from the underlying collective model and matching model. There is no need to assume that Pareto weights are given by some unknown function of prices, income and distribution factors. For example, it is clear that the wage ratio should affect the intra-household allocation of bargaining power in my model, because the wage ratio is explicitly related to the outside options of men and women. Changing the identity of the recipient of some nonlabor income (i.e a transfer that is no longer given to men, but to the women, as in the natural experiment studied in \textcite{LundbergPollakWales1997}) does change the allocation of bargaining power as well, providing this also affects the outside options of men and women. The same applies for the sex ratio, since changing population supplies changes the market structure and yields to a new equilibrium on the marriage market\footnote{This paragraph underlines the structural nature of my approach, because there is no reason for a distribution factor to affect the intra-household allocation of power unless it appears explicitly in the model.}.

\section{Conclusion}\label{sec:conclusion}

The present contribution brings together two important parts of the family economics literature: collective models, that describe the bargaining process that takes place within the household, and matching models, that have been used to model the marriage market.
This is an important step forward because collective models, although very powerful, are incomplete: both household formation and the allocation of power within the household are taken as given. Modelling the marriage market jointly with household bargaining is one way of dealing with these shortcomings.

There have been a few attempts to achieve this goal using matching models with perfectly transferable utility. While this approach has produced impressive results, it is not completely satisfying. In particular, ``power'' does not matter in determining the public good consumption of married individuals. Therefore, I have proposed in this paper to use the matching framework with imperfectly transferable utility developed in a prior work \parencite{GalichonKominersWeber2019} to close the gap between collective models and the marriage market. I tackle two issues that otherwise would prevent us from reaching this goal. The first is to characterize a class of collective models that produce proper bargaining sets, and therefore that can be embedded in the ITU framework. The second is to provide a way to estimate these models, which is more challenging than in a TU framework. I propose an approach based on MPEC and an efficient computational method to obtain distance functions. In addition, the method underlines the connections that exist between the ITU matching framework and collective models.

Finally, I provide a fully-fledged application using PSID data and a non-trivial model that includes private consumption, leisure and a public good that is produced from time inputs. I investigate whether a parsimonious model fit the data well. I show that the sharing rule has increased from 0.392 in 1969 to 0.435 in 2017. I illustrate the strength of the approach with counterfactual experiments. For example, I show that in 2017, were the gender wage gap to disappear on average, the sharing rule would increase to $0.514$, and the sharing of housework would decrease from $63\%$ to $55\%$.

The approach developed in this paper is very general, and could be used in conjunction, or compared to, other approaches such as \textcite{CherchyeDemuynckDeRockEtAl2017}. The model presented in the application is voluntarily parsimonious, but could easily be extended to endogenize fertility choices and time (and money) invested in children. Finally, beyond the collective models studied in this paper, the methods developed
here are applicable in other settings as well, e.g. on the labor market. These extensions are left for future research.

\pagebreak
\printbibliography[heading=none]

\pagebreak
\appendix
\section{Proofs}\label{app:proofs}

\subsection{Proof of Proposition \ref{prop:Fproper}}

\begin{proof}
The fact that $\mathcal{G}$ is lower comprehensive follows directly from its definition and the assumption of free disposal. Let us show that $\mathcal{G}$ is closed.
Consider a convergent sequence $\{w_{n}\}_{n\in\mathbb{N}}$ in $\mathcal{G}$
such that $w_{n}=(u_{n},v_{n})\in\mathcal{G}$, $\forall n\in\mathbb{N}$. For
each $w_{n}$, $\exists \omega_{n}\in\Omega:u_{n}\leq U(q^a_{n},Q_{n}),v_{n}\leq
V(q^b_{n},Q_{n}).$ Since $\Omega$ is compact, there exist a\ convergent
subsequence $\{\omega_{f(n)}\}_{n\in\mathbb{N}}$ in $\Omega$ (by the Bolzano–Weierstrass theorem). From
$\{\omega_{f(n)}\}_{n\in\mathbb{N}}$, construct the subsequence $\{w_{f(n)}%
\}_{n\in\mathbb{N}}$. For any $w_{k}$ in that subsequence, $\exists \omega_{k}%
\in\Omega:u_{k}\leq U(q^a_{k},Q_{k}),v_{k}\leq V(q^b_{k},Q_{k})$. Since
$u_{k}\rightarrow u$, $v_{k}\rightarrow v$ and $\omega_{k}\rightarrow
\omega=(q^a,q^b,Q)\in\Omega$, we see by continuity of $U$ and $V$ that $u\leq
U(q^a,Q)$ and $v\leq V(q^b,Q)$. Therefore $(u,v)\in\mathcal{G}.$ Point (iii) in the definition of a proper bargaining set follows immediately from the fact that $U$ and $V$ are bounded above on $\Omega$.
\end{proof}

\subsection{Proof of Proposition \ref{prop:weakPE}}
\begin{proof}
The proof follows immediately from the fact that if $\mathcal{G}$ is a proper bargaining set, it is lower comprehensive and closed. Indeed, take any $(u,v)$ on the boundary of $\mathcal{G}$. If $(u,v)$ is not weakly Pareto efficient, $\exists(u^{\prime},v^{\prime}%
)\in\mathcal{G}$ such that $u^{\prime}>u$ and $v^{\prime}>v$, hence $(u,v)$ is an interior point (consider the open ball of center $(u,v)$ and radius $\epsilon$ $B_{\epsilon}$ where $\epsilon<\min(u^{\prime}-u,v^{\prime}-v)$).
\end{proof}

\subsection{Proof of Proposition \ref{prop:strongPE}}
\begin{proof}
Choose any $(u,v)$ on the boundary of $\mathcal{H}$ attained with the
allocation $\omega = (q^a,q^b,Q)$ and suppose it is not Pareto efficient. Then, there is a
point $(u^{\prime},v^{\prime})\in\mathcal{H}$ attained for some allocation
$\omega^\prime = (q^{a^\prime},q^{b\prime},Q^{\prime})$ such that $u^{\prime}\geq u$ and
$v^{\prime}\geq v$, with at least one strict inequality. Without loss of
generality, say $u^{\prime}>u$. Note that $v^{\prime}>v$ yields to a
contradiction since $(u,v)$ would be an interior point (see the proof of Proposition \ref{prop:weakPE}). Hence, $u^{\prime}>u$ and
$v^{\prime}=v$. To reach $u^{\prime}$, we need to change the amount of goods consumed by the man, so that either $q^{a^\prime}\neq q^a$ or $Q^{\prime}\neq Q$, or both.

If $Q^{\prime}\neq Q$, then we can show that $(u,v)$ is an interior point, a contradiction. Indeed, the allocation
$(tq^a+(1-t)q^{a^\prime},tq^b+(1-t)q^{b\prime},tQ+(1-t)Q^\prime ) \in \Omega$ for $t \in (0,1)$ by convexity of $\Omega$,
and gives both partners utility  $(u^{\prime\prime},v^{\prime\prime})$ with $u^{\prime\prime}>u$ and $v^{\prime\prime}>v$.

If $Q^{\prime}= Q$, then $q^{a^\prime}\neq q^a$ and there is at least one private good $k$ for which
$q_{k}^{a^\prime}>q^a_{k}\geq0$ since $U$ is strictly increasing. By continuity of
$U$ and $V$, and Assumption \ref{ass:transfers}, we can always slightly
decrease $q_{k}^{a^\prime}$ and slightly increase $q^b_{l}$ for some private good
$l$ in a feasible way, and reach a point $(u^{\prime\prime},v^{\prime\prime})$
such that $u^{\prime}>u^{\prime\prime}>u$ and $v^{\prime\prime}>v$. Hence,
$(u,v)$ is an interior point, a contradiction.
\end{proof}

\subsection{Proof of Proposition \ref{prop:properfinal}}
\begin{proof}
Closedness and boundedness are proven exactly as in Proposition
\ref{prop:Fproper}. We need to show that $\mathcal{H}$ is lower comprehensive.
Take $(u,v)\in\mathcal{H}$, with $u=U(q^a,Q)$ and $v=V(q^b,Q)$ for some allocation
$(q^a,q^b,Q)\in\Omega$ and consider a point $(u^{\prime},v^{\prime})$ such
that $u^{\prime}\leq u,v^{\prime}\leq v$. Without loss of generality, take
$u^{\prime}<u$. \ By Assumption \ref{ass:UV} and
\ref{ass:limit}, we can find $q_{1}^{a^\prime}$ such that $((q_{1}^{a^\prime
},q^a_{-1}),q^b,Q)\in\Omega$, $U((q_{1}^{a^\prime},q^a_{-1}),Q)=u^{\prime}$ and
$V(q^b,Q)=v$.

If $v^{\prime}=v$, then we stop.

If $v^{\prime}<v$, there is a $q_{1}^{b\prime}$ such that $((q_{1}^{a^\prime
},q^a_{-1}),(q_{1}^{b\prime},q^b_{-1}),Q)\in\Omega$, $U((q_{1}^{a^\prime}%
,q^a_{-1}),Q)=u^{\prime}$ and $V((q_{1}^{b\prime},q^b_{-1}),Q)=v^{\prime}$.
Therefore, $(u^{\prime},v^{\prime})\in\mathcal{H}$.
\end{proof}

\subsection{Proof of Theorem \ref{thm:gradient}}
\begin{proof}
The proof follows from the Envelop Theorem, since
\[
D_{ij}^{\theta}(u,v) = \min_{z_{ij},q^a_{ij},q^b_{ij},Q_{ij}}z_{ij}\text{ \thinspace}s.t\text{
(\ref{eq:CompMethodii}).}
\]
\end{proof}

\subsection{Proof of Theorem \ref{thm:collectivemodels}}
\begin{proof}
To prove (i), we need to show that the point $(u-z_{ij}^\star, v-z_{ij}^\star)$ (where $z_{ij}^\star$ solves problem \ref{eq:P1} is a boundary point of  $\mathcal{H}_{ij}$, and therefore it is Pareto efficient by Proposition \ref{prop:strongPE}. If $(u-z_{ij}^\star, v-z_{ij}^\star)$ is not a boundary point of $\mathcal{H}_{ij}$, then there is an open ball of center $(u-z_{ij}^\star, v-z_{ij}^\star)$ and radius $\epsilon$ completely contained in $\mathcal{H}_{ij}$. Therefore, the point $(u-z_{ij}^\star+\frac{\epsilon}{2}, v-z_{ij}^\star+\frac{\epsilon}{2})$ is also in $\mathcal{H}_{ij}$, but then $z_{ij}^\star$ cannot be solution to problem \ref{eq:P1}.

To prove (ii), we write the Lagrangian associated to problem \ref{eq:P1}, except that we maximize $-z_{ij}$ instead of minimizing $z_{ij}$:
\begin{align*}
  \mathcal{L}(z_{ij},q^a_{ij},q^b_{ij},Q_{ij},\lambda_1,\lambda_2,\xi_r) = -z_{ij} &- \lambda_1 (u-z_{ij} -U_{ij}^{\theta}(q^a_{ij},Q_{ij})) \\
   & - \lambda_2 (u-z_{ij} -V_{ij}^{\theta}(q^b_{ij},Q_{ij})) \\
   & - \sum_{r}^{R} \xi_r g_{ij}^r(q^a_{ij},q^b_{ij},Q_{ij})
\end{align*}
For a solution point $(z_{ij}^\star,q_{ij}^{a\star},q_{ij}^{b\star},Q_{ij}^\star)$, I introduce the associated nonnegative multipliers $\lambda_1^\star, \lambda_2^\star, \xi_r^\star$. Consider now the following Lagrangian:
\[
\tilde{\mathcal{L}}(q^a_{ij},q^b_{ij},Q_{ij},\xi_r) = \lambda_1^\star U_{ij}^{\theta}(q^a_{ij},Q_{ij}) + \lambda_2^\star V_{ij}^{\theta}(q_{ij}^b,Q_{ij}) - \sum_{r}^{R} \xi_r g_{ij}^r(q^a_{ij},q^b_{ij},Q_{ij})
\]
and note that (i) $\tilde{\mathcal{L}}$ is a concave function of $(q^a_{ij},q^b_{ij},Q_{ij})$, (ii) taking $\xi_r=\xi_r^\star$ and

$(q^a_{ij},q^b_{ij},Q_{ij})=(q_{ij}^{a\star},q_{ij}^{b\star},Q_{ij}^\star)$, we have $\xi_r^\star\geq 0$ and
\[
\frac{\partial \mathcal{L}\tilde(q_{ij}^{a\star},q_{ij}^{b\star},Q_{ij}^\star, \xi_r^\star)}{\partial q_{ij}^{a}} = 0,
\frac{\partial \mathcal{L}\tilde(q_{ij}^{a\star},q_{ij}^{b\star},Q_{ij}^\star, \xi_r^\star)}{\partial q_{ij}^{b}} = 0, \text{ and }
\frac{\partial \mathcal{L}\tilde(q_{ij}^{a\star},q_{ij}^{b\star},Q_{ij}^\star, \xi_r^\star)}{\partial Q_{ij}} = 0
\]
Therefore, $(q_{ij}^{a\star},q_{ij}^{b\star},Q_{ij}^\star)$ is solution to
\begin{align*}
max_{q_{ij}^{a},q_{ij}^{b},Q_{ij}} \lambda_1^\star U_{ij}^{\theta}(q^a_{ij},Q_{ij}) + \lambda_2^\star V_{ij}^{\theta}(q^b_{ij},Q_{ij})\\
\text{s.t } g_{ij}^r(q_{ij}^{a},q_{ij}^{b},Q_{ij})  \leq0,\text{ }r\in\{1,...,R\}
\end{align*}
Hence, $(q_{ij}^{a\star},q_{ij}^{b\star},Q_{ij}^\star)$ is a Pareto efficient allocation that maximizes a social welfare function with Pareto weights $\lambda_1^\star$ and $\lambda_2^\star$. The fact that $\lambda_1^\star+\lambda_2^\star=1$ follows directly from the first order conditions of problem \ref{eq:P1}, since $\partial \mathcal{L}/\partial z_{ij} = 0 \iff -1 +\lambda_1+\lambda_2 = 0$.
\end{proof}

\section{Computation method for general proper bargaining sets}\label{app:comp}

\begin{problem}\label{eq:P2}
Suppose that $\mathcal{G}_{ij}$ is a proper bargaining set associated with some collective model with free disposal, with preferences and feasible set of goods $(U_{ij}^{\theta}, V_{ij}^{\theta}, \Omega_{ij})$. Then we can solve
 \begin{align}
&  \min_{z_{ij},q^a_{ij},q^b_{ij},Q_{ij}}z_{ij}\\%
\begin{split}
s.t\text{ } u-z_{ij}  &  \leq\mathcal{U}_{ij}^{\theta}(q^a_{ij},Q_{ij})\\
            v-z_{ij}  &  \leq\mathcal{V}_{ij}^{\theta}(q^b_{ij},Q_{ij})\\
            g_{ij}^r(q^a_{ij},q^b_{ij},Q_{ij})  &  \leq0,\text{ }r\in\{1,...,R\}
\end{split}
\end{align}
when $D^{\theta}_{ij}(u,v) = z_{ij}^*$, solution to the above program.
\end{problem}

The solution method is the same as in the main text, except for the inequality constraints $u-z_{ij} \leq\mathcal{U}_{ij}^{\theta}(q^a_{ij},Q_{ij})$ and
$v-z_{ij}  \leq\mathcal{V}_{ij}^{\theta}(q^b_{ij},Q_{ij})$.

\section{Log-likelihood estimation}\label{app:estim}
The construction of the log-likelihood follows the same step as in GKW. Our model predicts leisure and housework
time for any $(i,j)$ pair of individuals and for singles. I use this information and the assumption that housework and leisure are observed in the data with some Gaussian measurement error (with variance $s_1^2$, $s_2^2$, $s_3^2$, and $s_4^2$ for the labor supply of men and women and the housework of men and women, respectively) to construct the likelihood function. Since I assumed a uniform distribution of types, we may as well reintroduce the notation $(i,j)$ instead of types. We can form the log-likelihood as follows:
\begin{eqnarray*}
\log \mathcal{L}(\theta ,u_{i}^{\theta },v_{j}^{\theta }) &=&
-\sum_{(i,j)\in \mathcal{C}}\left[
D^{\theta }(u_{i}^{\theta },v_{j}^{\theta })+
\frac{1}{2}\left( \frac{\ell^a_{ij}-\hat{\ell}^a_{ij}}{{s}_{1}}\right) ^{2}+
\frac{1}{2}\left( \frac{\ell^b_{ij}-\hat{\ell}^b_{ij}}{{s}_{2}}\right) ^{2}\right] \\
&&-\sum_{(i,j)\in \mathcal{C}}\left[\frac{1}{2}\left( \frac{h^a_{ij}-%
\hat{h}^a_{ij}}{{s}_{3}}\right) ^{2}+\frac{1}{2}\left( \frac{h^b_{ij}-\hat{h}^b%
_{ij}}{{s}_{4}}\right) ^{2}\right]  \\
&&-\sum_{i\in \mathcal{S}_{M}}\left[ u_{i}^{\theta }+
\frac{1}{2}\left( \frac{\ell^s _{i}-\hat{\ell}^s_{i}}{{s}_{1}}\right) ^{2} +
\frac{1}{2}\left( \frac{h^s _{i}-\hat{h}^s_{i}}{{s}_{3}}\right) ^{2} \right]  \\
&&-\sum_{j\in \mathcal{S}_{F}}\left[ v_{j}^{\theta }+\frac{1}{2}\left( \frac{\ell^s _{j}-%
\hat{\ell}^s_{j}}{{s}_{2}}\right) ^{2} +\frac{1}{2}\left( \frac{h^s _{j}-%
\hat{h}^s_{j}}{{s}_{4}}\right) ^{2} \right]  \\
&&-|\mathcal{I}|\log {s}_{1}-|\mathcal{J}|\log {s}_{2}-|\mathcal{C}|\log {s}_{3}-|\mathcal{C}|\log {s%
}_{4} \\
&&-\hat{N}\log(N)
\end{eqnarray*}
In the above expression, $\ell^a_{ij}$, $\ell^b_{ij}$, $h^a_{ij}$ and $h^b_{ij}$ ($\ell^s_{i}$, $\ell^s_{j}$, $h^s_{i}$ and $h^s_{j}$) denote the predicted leisure and housework time of (single) married men and women, respectively. The observed counterparts are denoted $\hat{\ell}^a_{ij}$, $\hat{\ell}^b_{ij}$, $\hat{h}^a_{ij}$ and $\hat{h}^b_{ij}$ ($\hat{\ell}^s_{i}$, $\hat{\ell}^s_{j}$, $\hat{h}^a_{i}$ and $\hat{h}^s_{j}$ for singles). $\hat{N}$ and $N$ are the observed and predicted number of households. The set of matched pairs $(i,j)$ observed in the data is denoted  $\mathcal{C}$, and $\mathcal{S}_{M}$ and $\mathcal{S}_{F}$ denote the set of single men and the set of single women observed in the data, respectively. The log-likelihood is maximized over $(\theta,\{u_i\}_{i\in\mathcal{I}},\{v_j\}_{j\in\mathcal{J}})$ subject to the constraint that the marriage market is in equilibrium (MPEC approach). Note that it is not necessary to optimize over $(s_1,s_2,s_3,s_4)$ as it is possible to solve directly for their optimal value. For example, $s_1 = \frac{1}{|\mathcal{I}|} \left(\sum_{(i,j)\in \mathcal{C}} (\ell^a_{ij}-\hat{\ell}^a_{ij})^2 + \sum_{i\in \mathcal{S}_{M}} (\ell^s _{i}-\hat{\ell}^s_{i})^2\right)$.

\section{Additional results}\label{app:results}
\subsection{Observed matching by education}

\FloatBarrier
\begin{table}[h]
  \centering
  \caption{Observed matching, by education}\label{tab:matching}
  \begin{tabular}{lM{3cm}M{3cm}M{3cm}}
  \toprule
  & Non-college & College & $\{0\}$ (singlehood)\\
  \midrule
   \textbf{A. 1977.} & &   & \\
   Non-college &142 &31 &17 \\ 
College &45 &44 &9 \\ 
$\{0\}$ (singlehood) &43 &14 &
\\

   &&&\\
   \textbf{B. 1997.} & &   & \\
   Non-college &105 &53 &40 \\ 
College &67 &233 &73 \\ 
$\{0\}$ (singlehood) &65 &107 &
\\

   &&&\\
   \textbf{C. 2017.} & &   & \\
   Non-college &47 &64 &50 \\ 
College &24 &188 &60 \\ 
$\{0\}$ (singlehood) &47 &117 &
\\

   \bottomrule
  \end{tabular}
  \caption*{\raggedright\tiny This table shows the observed matching, by education level, for three selected marriage markets: 1977, 1997 and 2017.}
 \end{table}
\FloatBarrier

\subsection{Model fit}
\FloatBarrier
\begin{table}[h]
  \centering
  \caption{Model Fit}\label{tab:modelfit}
  \begin{tabular}{lM{2cm}M{2cm}M{2cm}M{2cm}}
  \toprule
  & $T-h^a_{xy}-\ell^a_{xy}$ & $h^a_{xy}$ & $T-h^b_{xy}-\ell^b_{xy}$ & $h^b_{xy}$\\
  \midrule
   \textbf{A. 1977.} & &  &  & \\
   Observed (mean) &41.498 &6.021 &18.378 &28.567 \\ 
Predicted (mean) &41.414 &4.448 &17.069 &28.339 \\

   &&&&\\
   \textbf{B. 1997.} & &  &  & \\
   Observed (mean) &42.340 &6.638 &30.223 &16.948 \\ 
Predicted (mean) &41.779 &5.187 &29.665 &16.469 \\

   &&&&\\
   \textbf{C. 2017.} & &  &  & \\
   Observed (mean) &40.531 &8.209 &32.210 &13.941 \\ 
Predicted (mean) &40.109 &7.141 &32.220 &13.034 \\ 
\\

   \bottomrule
  \end{tabular}
  \caption*{\raggedright\tiny This table shows the fit of the model for three selected marriage markets: 1977, 1997 and 2017. The first and second rows for each market display the mean of the observed and predicted dependent variables.}
 \end{table}

\end{document}